\documentclass{osa-article}

\journal{ao}

\articletype{Research Article}

\usepackage{graphicx}
\usepackage{subcaption}
\usepackage{wrapfig}

\begin{document}

\title{Beam propagation simulation of phased laser
arrays with atmospheric perturbations}

\author{Will Hettel,\authormark{1} Peter Meinhold,\authormark{1} Jonathan Y. Suen,\authormark{1} Prashant Srinivasan,\authormark{1} Peter Krogen,\authormark{1} Allan Wirth,\authormark{2} and Philip Lubin\authormark{1}}

\address{\authormark{1}Physics Department, University of California, Santa Barbara, CA 93106, USA}
\address{\authormark{2}Massachusetts Institute of Technology Lincoln Laboratory, Lexington, MA 02421-6426, USA}

\email{whettel@physics.ucsb.edu} 



\begin{abstract}
Directed energy phased array (DEPA) systems have been proposed for novel applications such as beaming optical power for electrical use on remote sensors, rovers, spacecraft and future moon bases, as well as planetary defense against asteroids and photonic propulsion up to relativistic speeds. All such scenarios involve transmission through atmosphere and beam perturbations due to turbulence which must be quantified. Numerical beam propagation and feedback control simulations were performed using an algorithm optimized for efficient calculation of real-time beam dynamics in a Kolmogorov atmosphere. Results were used to quantify the effectiveness of the system design with different degrees of atmospheric turbulence and zenith angles, and it was found that a large aperture DEPA system placed at a high altitude site is capable of producing a stable diffraction limited spot (Strehl > 0.8) on space-based targets for Fried length $r_0\geq10$~cm (@500nm) and zenith angles up to 60 degrees depending on atmospheric conditions. These results are promising for the next generation of power beaming and deep space exploration applications.
\end{abstract}

\section{Introduction}
Directed energy phased array (DEPA) systems are preferable to single aperture systems in directed energy applications that require large apertures for long range targets because of their scalability in both power and size. Scalability in power opens up a new mission space with high power requirements, and scalabilty in size allows illumination of targets at greater distances with minimal loss from diffraction. Existing monolithic single mode laser elements are limited to $\sim$10~kW in power \cite{Zervas2014_hplasers} and meter-class aperture sizes at maximum. In principle, a DEPA could be built to have any power or diameter by incorporating the appropriate number of modular subelements. Furthermore, the ability to control the phase and alignment of each laser subelement provides a means to correct for atmospheric turbulence. These features make DEPAs a potential solution to various problems involving energy supply to distant objects on the ground, in the air or in space in scenarios where sunlight is insufficient or inaccessible. For example, sending power to the moon during the lunar night could be accomplished by using an Earth-based DEPA to provide optical power to a high efficiency tuned  photovoltaic array on the moon. Power could also be sent to other extraterrestrial targets such as low Earth orbit, medium Earth orbit, geosynchronous and deep space spacecraft for power or propulsion. The latter could be done via photovoltaic conversion on the spacecraft to power high specific impulse ion engines and enable station keeping or rapid interplanetary transit. Kilometer scale arrays would be capable of diverting incoming asteroids off course by ablation, or accelerating low mass spacecraft to relativistic speeds for interstellar missions \cite{Lubin2016_roadmap,Lubin2020_ESA}. Other applications include but are not limited to deep space communications, powering of ground-based remote sensors and rovers in inhospitable locations, remote sensing LIDAR, laser machining, and experiments involving matter under extreme conditions.

\begin{figure}
    \centering
    \includegraphics[width=1\textwidth]{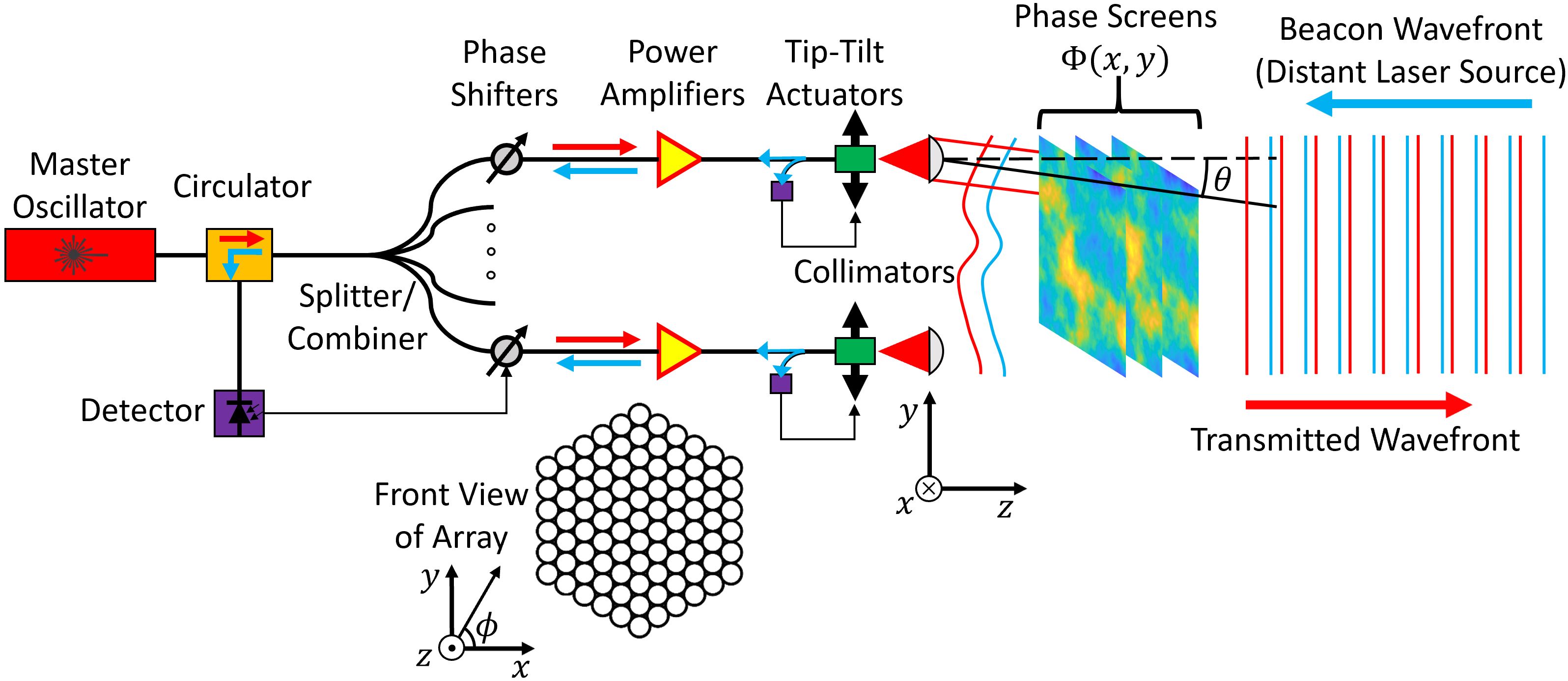}
    \caption{Schematic diagram of the simulated system with coordinates. A beacon laser signal (right, blue) passes through atmospheric phase screens and illuminates the DEPA (left), which measures the reverse-propagating beacon in fiber for phase conjugation and alignment feedback. The transmit beam (red) originates from a master seed laser, which is split and amplified in fiber before propagating in free space. Here a 91 element hexagonal array is shown, but the DEPA can be scaled to any size or shape.}
\label{fig:schematic}
\end{figure}

While it is feasible to construct DEPA systems for space-based applications in space to avoid atmospheric effects and increase target visibility, cost and accessibility make it far more practical to start with ground based systems. It is therefore of great importance to quantify the systems' effectiveness in correcting for atmospheric perturbations. This study considers a ground-based continuous wave DEPA supplying power to a space-based target. This baseline DEPA is composed of a large number of circular sub-apertures of 10~cm diameter, which are hexagonally close packed so their edges touch each other. The target is equipped with a ``beacon'' laser, which illuminates the DEPA after being perturbed by the atmosphere. The atmosphere is modeled as a large number of two dimensional phase screens at different altitudes that represent the wavefront (phase) errors that a plane wave acquires as it travels in the zenith direction. The DEPA is equipped with sensors to measure the reverse-propagating beacon light and use it for phase retrieval and tip-tilt alignment. For simplicity we assume the target is stationary, making the results applicable to slow moving assets but not necessarily fast moving targets such as low Earth orbit satellites. Fig. \ref{fig:schematic} schematically represents the full system being simulated. The wavelength of both the beacon and transmit beams is assumed to be 1.064~$\mu$m to take advantage of developed ytterbium fiber laser amplification technology and low atmospheric absorption \cite{GeminiObs_transmission,Lubin2020_ESA}, though the model can easily be adapted to other wavelengths such as 1.55~$\mu$m (erbium-doped fiber amplifiers) and 2~$\mu$m (thulium-doped fiber amplifiers). This paper outlines how noise in such a system is modeled and provides simulated results for performance under various probable conditions at astronomical sites.

We utilize a recently developed DEPA beam simulation tool described in \cite{Hettel2019_spie} to calculate the transmitted beam profile. Conventional propagation methods, such as the fast Fourier transform based angular spectrum method \cite{Novotny2012_prop}, are not well suited to handle large numbers of array elements with uniquely defined perturbations on each subelement. The computation time and memory requirements for parameterizing the near field DEPA beam to sufficient resolution are significant, and any FFT based algorithm is poorly scalable with array size unless many spatial measurement points of the beam profile at the target are needed. Our method works by summing the analytic electric field solutions of each subelement at a given measurement point. This operation makes parameterizing the near field beam trivial, and for $N$ subelements with a fixed number of measurement points it requires $\mathcal{O}(N)$ calculations. While this study deals with a modest number of subelements, the incorporation of temporal dynamics makes scalability of critical importance to enable swift calculation of the beam profile at many points in time. We use this infrastructure to simulate extremely large numbers of elements ($N>10^{10}$, limited only by availability of computing resources) to calculate the beam profile at multiple points in time in parallel.

\section{Simulation pipeline}
\begin{figure}
    \centering
    \includegraphics[width=1\textwidth]{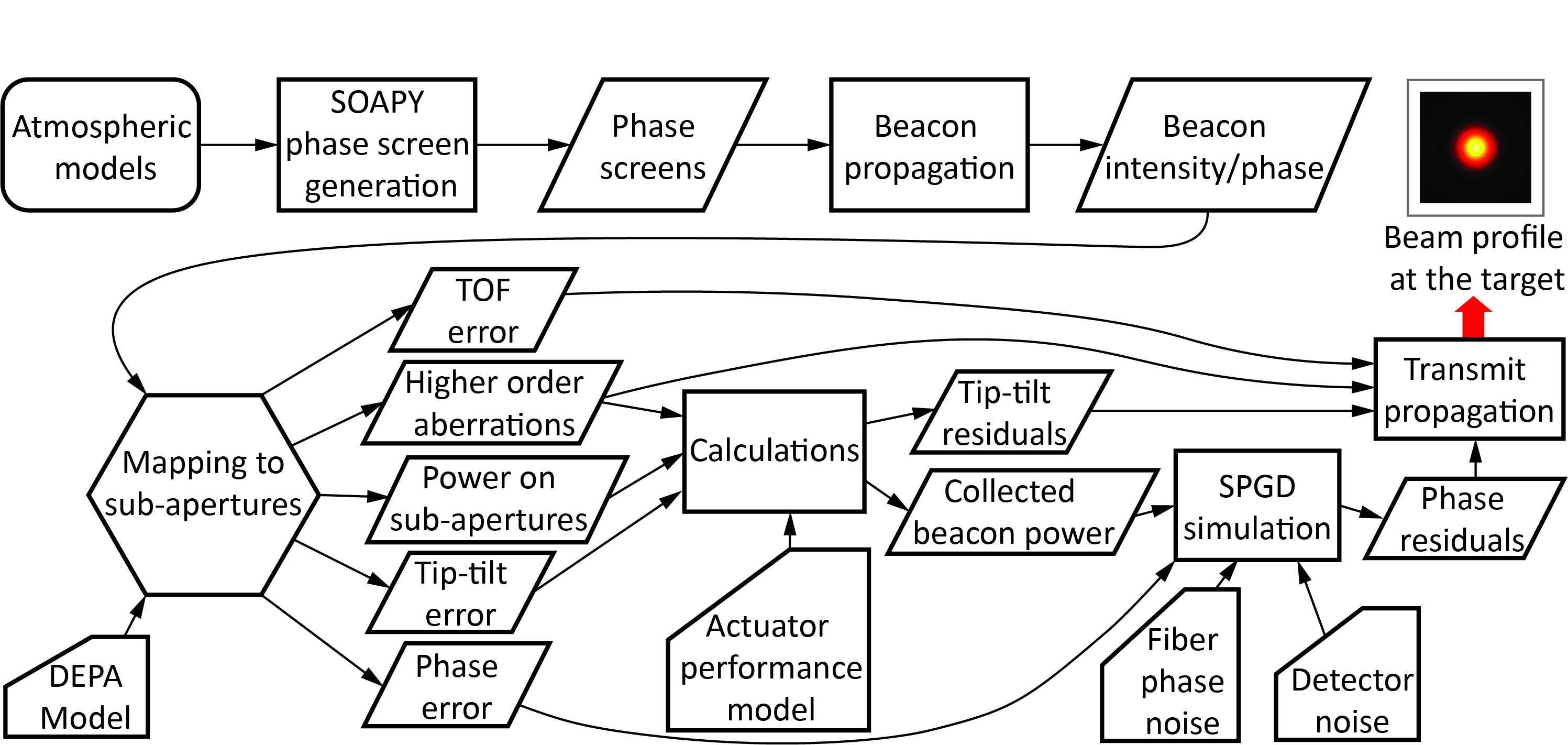}
    \caption{Simulation flow diagram including the various processes and noise inputs.}
\label{fig:pipeline}
\end{figure}

Fig. \ref{fig:pipeline} summarizes the steps of a DEPA atmospheric simulation and how they connect to each other. First, atmospheric phase screens are generated with the Python extension SOAPY \cite{Reeves2016_soapy} using a modified Hufnagel-Valley (HV) model for atmospheric turbulence, and a modified Bufton model for wind speed (see \textcolor{urlblue}{Supplement 1}, Section 2). The modifications of both models are informed by site survey data taken for the Thirty Meter Telescope (TMT) \cite{Schock2009_tmtdata}. The phase screens are used to optically propagate the beacon to the DEPA and produce its intensity and phase as a function of space and time. The data is then mapped to the locations of the DEPA sub-apertures to produce time series for piston phase, tip-tilt error, incident beacon power, intensity reduction due to higher order aberrations, and time-of-flight error for each subelement. A model for tip-tilt servo performance is used to generate tip-tilt residual error, which is combined with incident beacon power and higher order aberrations to calculate the beacon power collected by each subelement. Beacon power and piston phase values, along with detector noise and data-based fiber amplifier phase noise, are used as inputs for simulation of the stochastic parallel gradient descent (SPGD)  \cite{cbc2013_spgd} phasing servomechanism. SPGD is an algorithm routinely used in adaptive optics systems to phase lock numerous subelements. The SPGD simulation generates residual piston phase errors for the transmit beam. The phase residuals, tip-tilt residuals, intensity reduction due to higher order aberrations, and time-of-flight effects are used as inputs for the transmit beam propagator to produce final results for the beam at the target. Each step of the process is described in further detail in \textcolor{urlblue}{Supplement 1}, with exception of the optical propagation steps in the following section.

\section{Optical propagation of beacon and transmit beam}
\label{sec:propagation}
\begin{figure}
    \centering
        \begin{subfigure}[b]{0.45\linewidth}
        \includegraphics[width=\textwidth]{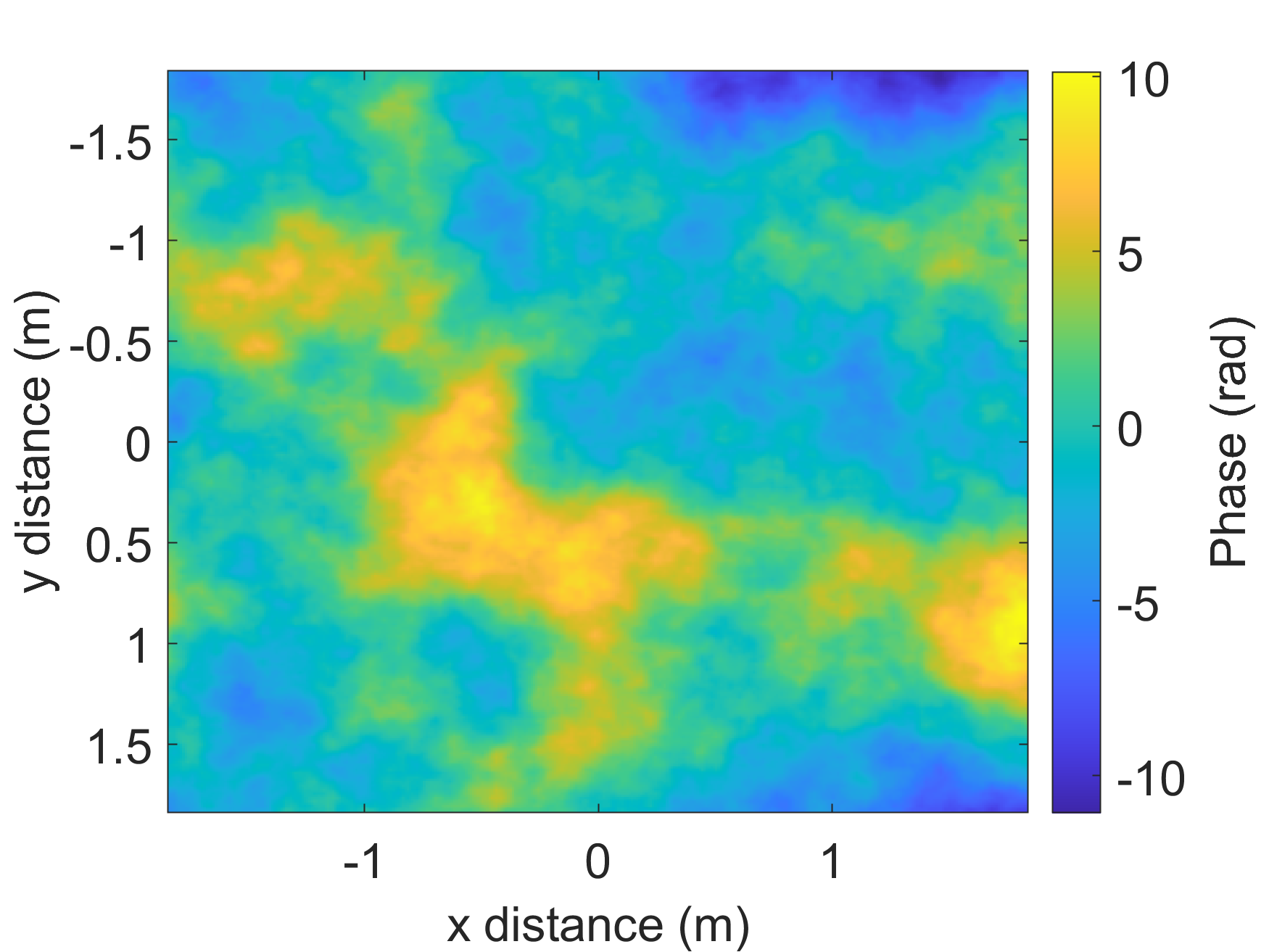}
        \caption{Simulated phase of beacon as it reaches the DEPA.}
        \label{fig:phasescreen}
    \end{subfigure}
    ~
    \begin{subfigure}[b]{0.45\linewidth}
        \includegraphics[width=\linewidth]{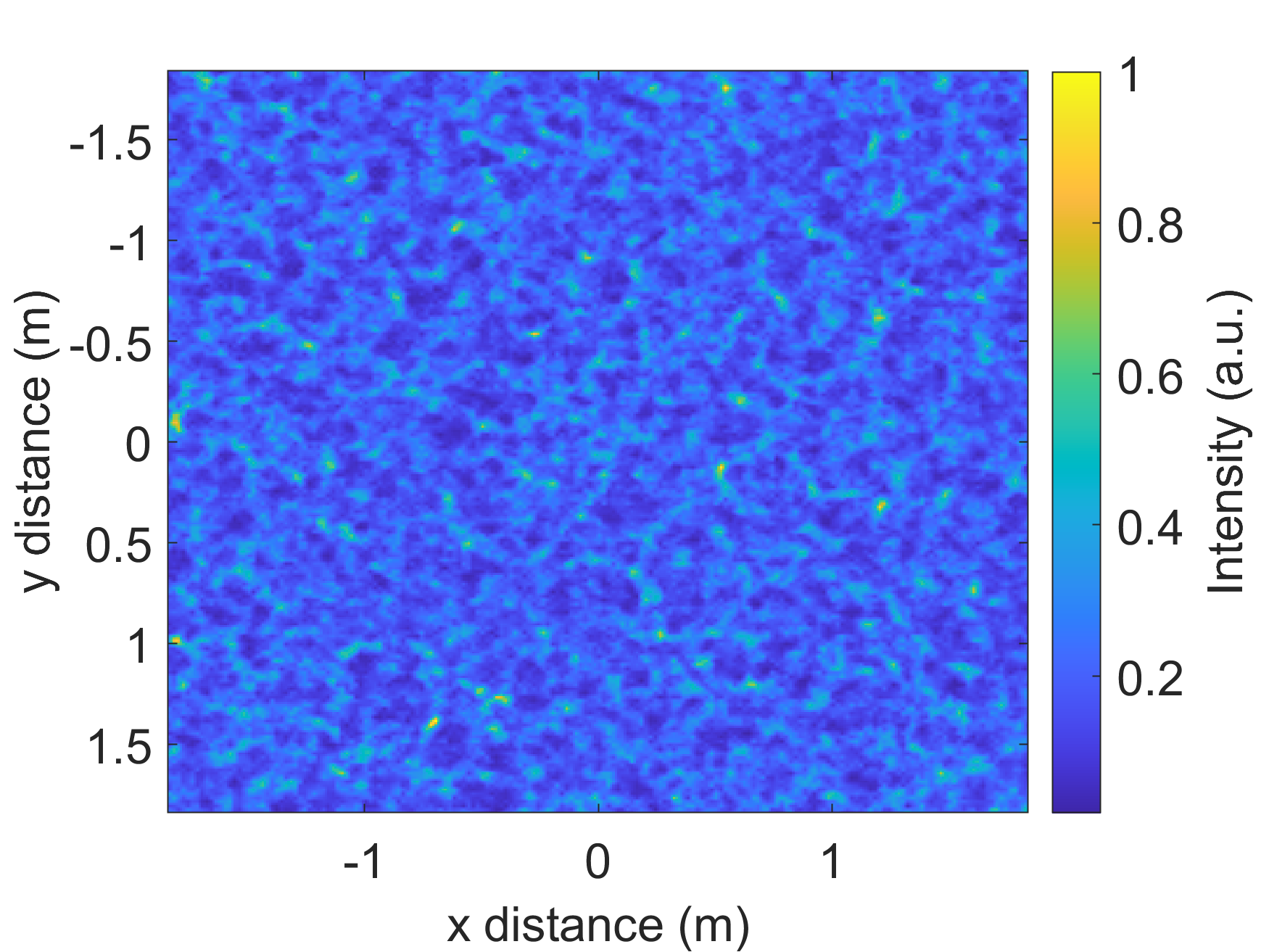}
        \caption{Simulated intensity of beacon as it reaches the DEPA.}
        \label{fig:beaconprofile}
    \end{subfigure}

    \caption{Simulated spatial profiles of phase (Fig. \ref{fig:phasescreen}) and intensity (Fig. \ref{fig:beaconprofile}) of a plane wave at 1.064 $\mu$m wavelength after propagating to the DEPA through an atmosphere with $r_0$ = 10~cm (for 500~nm wavelength) defined by the turbulence model discussed in \textcolor{urlblue}{Supplement 1}.}
\label{fig:beaconprop}
\end{figure}

The beacon is modeled as a plane wave propagating from the highest altitude phase screen to the ground. The electric field is iteratively calculated at the location of each phase screen to account for scintillation effects. Scintillation arises as an effect of the atmosphere refracting or scattering light in different directions, and the light interfering with itself as it propagates. The scintillation index $\sigma_I^2$ represents the normalized intensity variance with respect to time at a given point in space, and is defined as \cite{Andrews2001_scint}:
\begin{equation}
    \sigma_I^2 = \frac{<I^2>}{<I>^2} - 1.
\end{equation}
The plane wave model of the beacon makes initial parameterization of its electric field $\mathbf{E}_\mathrm{B}$ trivial, and we require a large number of spatial measurement points to accurately map it to the DEPA sub-apertures. These features make the angular spectrum method an appropriate propagator for $\mathbf{E}_\mathrm{B}$ \cite{Novotny2012_prop}:
\begin{equation}
\label{eq:angularspec}
\mathbf{E}_\mathrm{B}(x,y,z) = \iint_\infty^\infty \hat{\mathbf{E}}_\mathrm{B}(k_x,k_y;0)e^{i[k_xx + k_yy +k_zz]}dk_xdk_y
\end{equation}
where
\begin{eqnarray}
\hat{\mathbf{E}}_\mathrm{B}(k_x,k_y;z) &=& \frac{1}{4\pi^2}\iint_\infty^\infty \mathbf{E}_\mathrm{B}(x,y,z)e^{i[k_xx + k_yy]}dxdy \\
k_z &=& \sqrt{k^2 - k_x^2 - k_y^2} \ \ \ \approx \ \ \ k - \frac{k_x^2 + k_y^2}{2k}.
\label{eq:k}
\end{eqnarray}
Here $x$, $y$, and $z$ are the spatial coordinates in Fig. \ref{fig:schematic}, $\hat{\mathbf{E}}_\mathrm{B}$ is the Fourier transform (or angular spectrum) of $\mathbf{E}_\mathrm{B}$, and $\boldsymbol{k}$ is the wave vector of the propagating light. $\hat{\mathbf{E}}_\mathrm{B}$ is propagated in the $z$ direction by the $k_zz$ term in Eq. (\ref{eq:angularspec}), and inverse Fourier transformed to solve for $\mathbf{E}_\mathrm{B}$. The paraxial approximation is employed in Eq. (\ref{eq:k}). Both $\hat{\mathbf{E}}_\mathrm{B}$ and $\mathbf{E}_\mathrm{B}$ are efficiently calculated with fast Fourier transforms. Fig. \ref{fig:beaconprop} is an example of the phase and intensity of a beacon of 1.064 $\mu$m wavelength at a given point in time after it propagates to the DEPA through an atmosphere with Fried length $r_0$ = 10~cm (for 500~nm wavelength) defined by the turbulence model discussed in \textcolor{urlblue}{Supplement 1}. All values of $r_0$ in this paper unless otherwise specified are at a wavelength of 500 nm. $\sigma_I^2 \approx0.2$, in agreement with theoretical and experimental results \cite{Thomas2005_scint,Schock2009_tmtdata}.

The transmit beam of the DEPA is propagated using the method described in \cite{Hettel2019_spie}, which models each laser subelement as a Gaussian beam truncated by a circular aperture. The electric fields $\mathbf{E}_i$ of $N$ subelements are found with their respective analytic solutions and added together to construct the synthesized electric field $\mathbf{E}_\mathrm{T}$:
\begin{equation}
\label{eq:transmitfield}
\mathbf{E}_\mathrm{T}(x,y,z) = \sum_{i=1}^N \mathbf{E}_i(x,y,z).
\end{equation}
This method enables rapid parameterization of large DEPAs (above $10^{10}$ subelements demonstrated in \cite{Hettel2019_spie}) with perturbations in phase, amplitude, position, output angle, and focal distance. Customized parallel computing allows for their electric fields to be quickly calculated on a GPU to great accuracy. In this study, we used a GeForce GTX 1070 and Titan-Z desktop GPU with double point precision.

The atmosphere is approximated to be ``thin'' when calculating the transmit beam profile, i.e. the beam encounters all atmospheric turbulence in the near field before propagating into the far field or focal plane. This simplification is valid for the discussed simulations since the total propagation distance is much larger than the thickness of the atmosphere \cite{Brooker1985_thinatm,laserprop2005_thinatm}. Another  notable simplification is that the propagator does not model subelement aberrations of higher order than focal shifts. Losses in the main beam at the target due to these aberrations are still accounted for by reduction of the on-axis intensity of the subelements by the appropriate amount (see \textcolor{urlblue}{Supplement 1}). Since all transmit beam calculations are done in the far field of the array, any differing effects will be contained in side lobes far outside our region of interest.

\section{Numerical results for transmit beam}

\begin{figure}
    \centering
        \begin{subfigure}[b]{0.45\linewidth}
        \includegraphics[width=\textwidth]{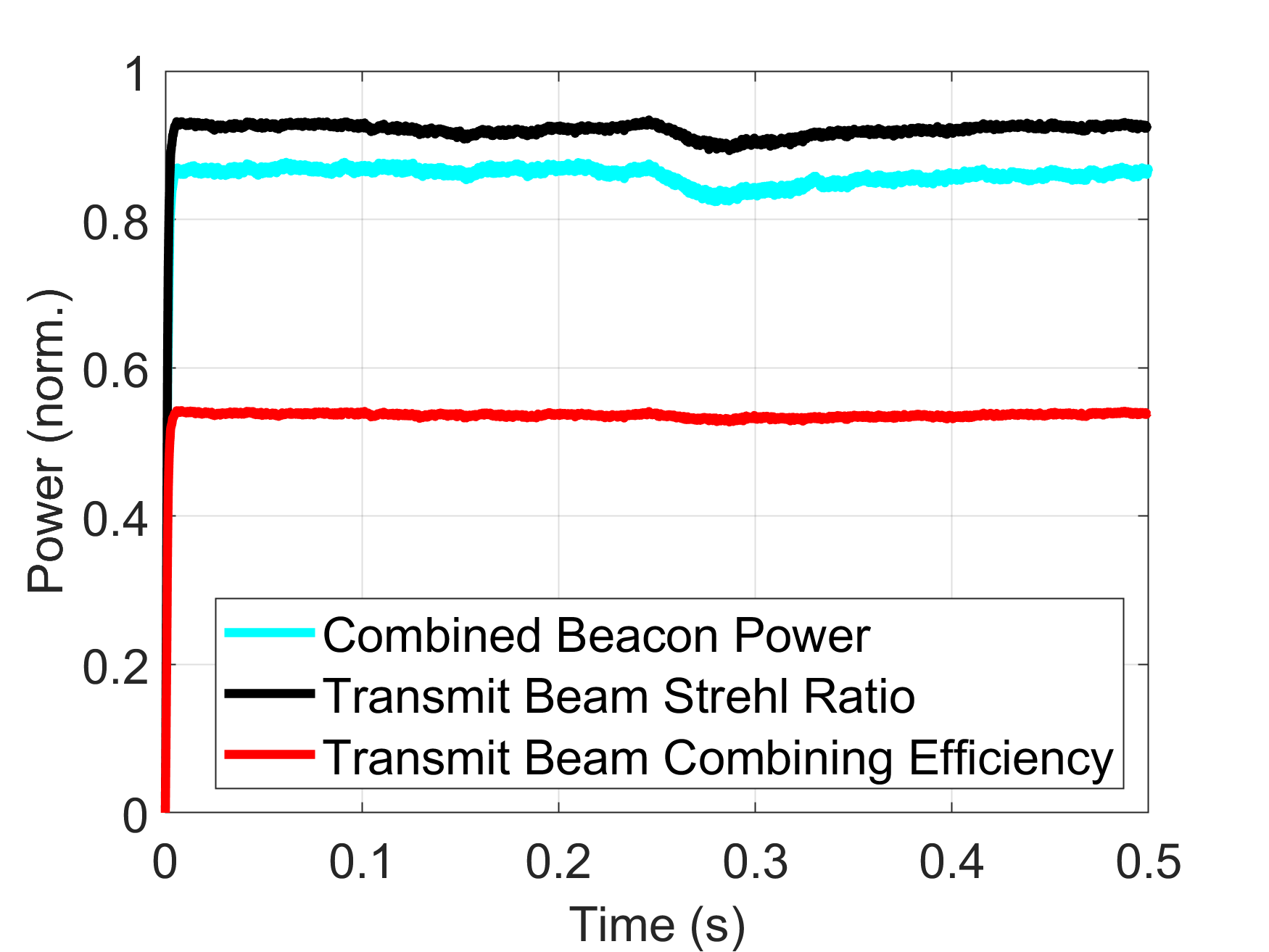}
        \caption{Time series of simulated measured beacon power, Strehl ratio of transmit beam, and combining efficiency of transmit beam.}
        \label{fig:timeseries}
    \end{subfigure}
    ~
    \begin{subfigure}[b]{0.45\linewidth}
        \includegraphics[width=\linewidth]{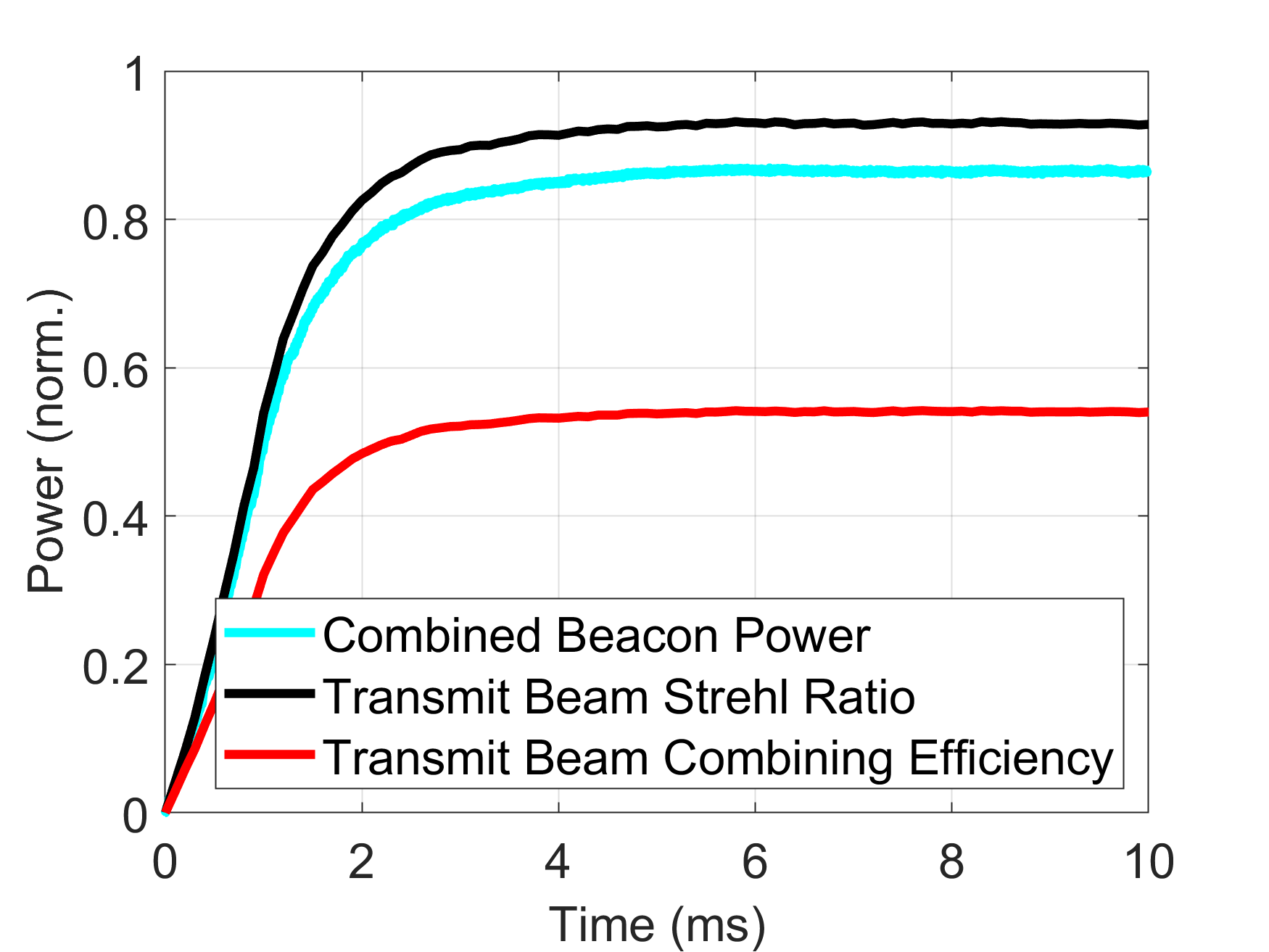}
        \caption{Same as left, but zoomed in on the phase convergence. The rise time is $\sim$2~ms, or about two cycles through the dither matrix as predicted for the SPGD algorithm in \textcolor{urlblue}{Supplement 1}.}
        \label{fig:timeserieszoom}
    \end{subfigure}

    \caption{Simulation output time series with $r_0$ = 15~cm (at 500~nm wavelength), $v_\mathrm{g}$ = 5~m/s and $\delta$ = 0~deg. Fig. \ref{fig:timeseries} shows the entire duration of the simulation, while Fig. \ref{fig:timeserieszoom} is zoomed in on the initial phase convergence. The beacon power is normalized to the amount that would be in the system with perfect alignment and no atmosphere. One can see that combined beacon power, output Strehl, and combining efficiency are well correlated with one another. With Strehl > 0.8, the beam is considered diffraction limited by the Maréchal criterion \cite{Born1975_marechal}.}
\label{fig:outputs}
\end{figure}

Simulations were performed for an array composed of $N=1024$ circular sub-apertures packed hexagonally with touching edges to form a disk. The sub-apertures were 10~cm in diameter, making the full array about 3.4~m in diameter. The transmit and beacon wavelengths were both 1.064~$\mu$m. While this is a specific configuration, the modularity of the subelements and known scaling relationships make the results applicable to a DEPA of any size (with the exception of piston phase errors) and targets at any distance. The measurement plane was set to be $4\times10^8$~m away from the array (approximately the distance to the moon) and was 500x500~m with 5~m resolution, which provides about 1\% error in estimating combining efficiency. Combining efficiency is defined as the ratio of optical power within an on-axis diffraction limited spot to the total power emitted by the array, and is a measure of the fractional power on the target. The spot diameter in angular units for a circular array of diameter $D$ and wavelength $\lambda$ is $2.44\lambda/D$, the first minimum of the Airy disk. For an array composed of circular sub-apertures which collimate Gaussian beams, the maximum achievable combining efficiency is $\sim$0.62 \cite{Hettel2019_spie,cbc2013_combining}. This limit is due to a combination of diffraction effects and power clipping at the apertures.

\begin{figure}
    \centering
    \begin{subfigure}[b]{0.45\linewidth}
        \includegraphics[width=\linewidth]{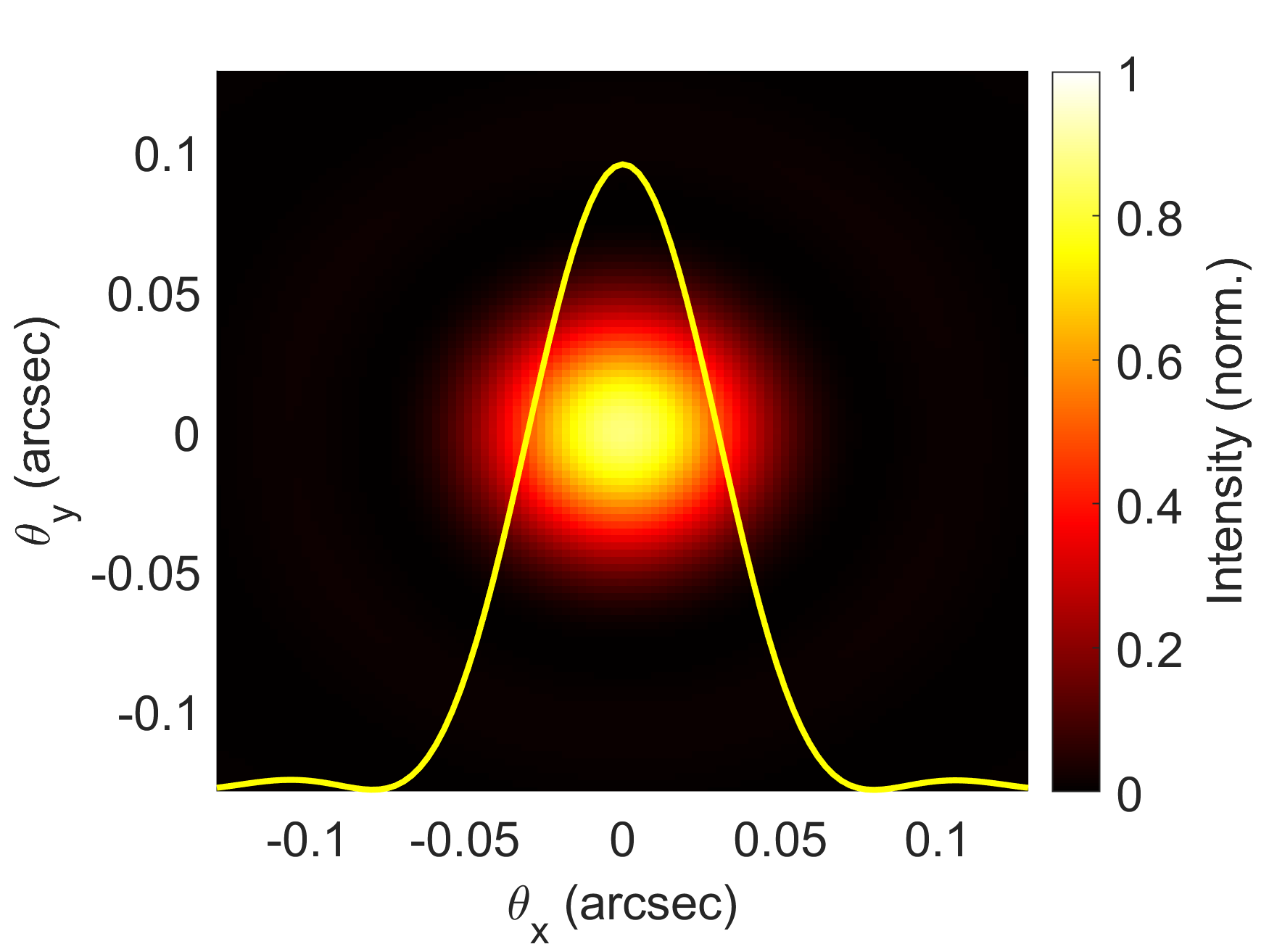}
        \caption{Simulated time averaged intensity of transmit beam in the target plane.}
        \label{fig:trasmitprofile}
    \end{subfigure}
    ~
    \begin{subfigure}[b]{0.45\linewidth}
        \includegraphics[width=\linewidth]{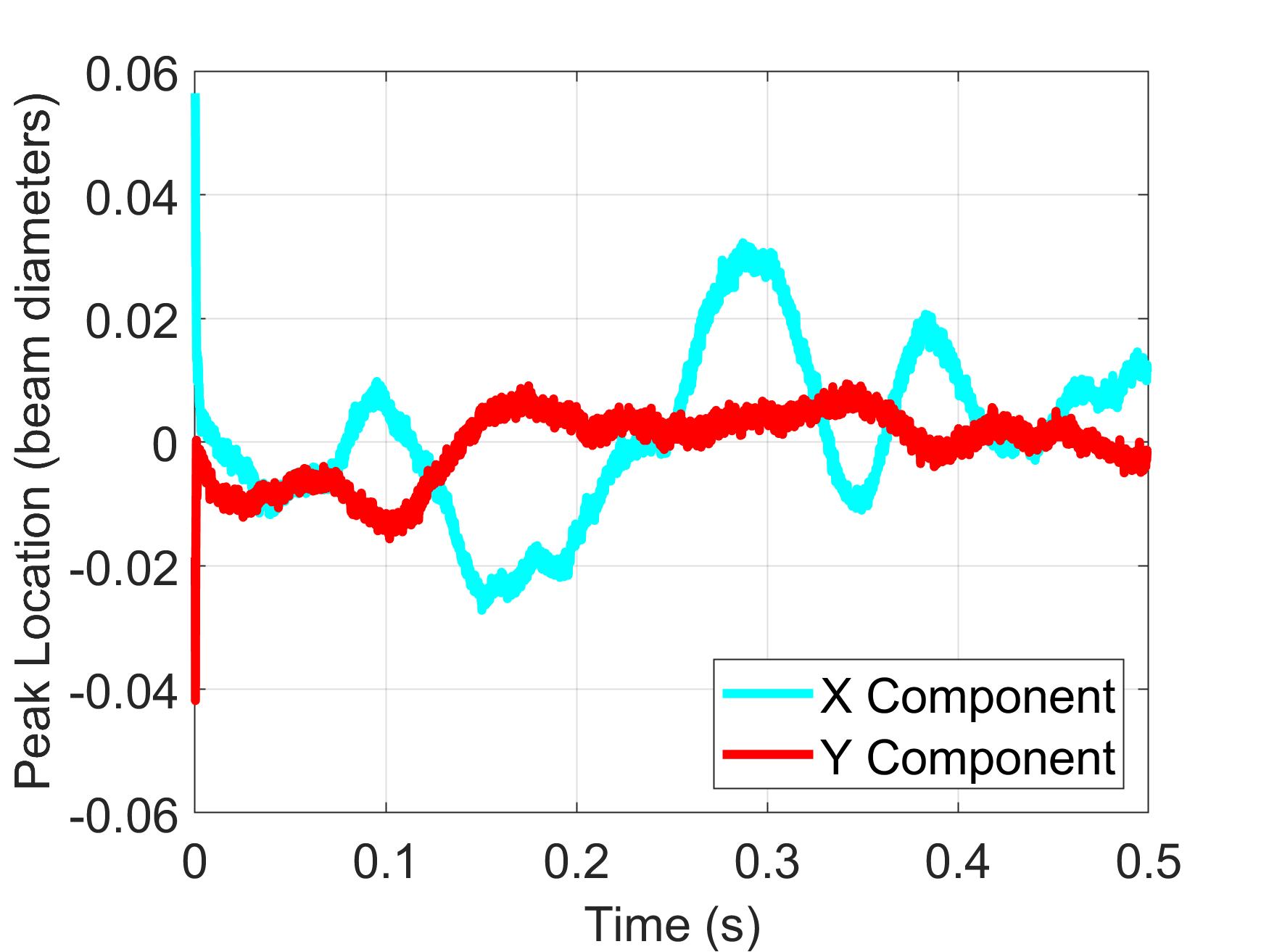}
        \caption{Peak location of simulated beam as a function of time.}
        \label{fig:peaklocation}
    \end{subfigure}

    \caption{Fig. \ref{fig:trasmitprofile} is a time average of the transmitted beam profile, normalized to its peak intensity with perfect transmission ($T$ = 1) and Strehl = 1. The color bar also functions as the y axis for the line plot. Fig. \ref{fig:peaklocation} shows the location of the peak as a function of time, normalized to the beam diameter as measured to first null. The wind is in the $x$ direction which causes larger deviations in $x$ than in $y$. Both plots are for $r_0$ = 15~cm, $v_\mathrm{g}$ = 5~m/s and $\delta$ = 0.} 
\label{fig:spatialprofile}
\end{figure}

Fig. \ref{fig:outputs} shows measured beacon power, the Strehl ratio of the transmit beam, and combining efficiency of the transmit beam for Fried length $r_0$ = 15~cm (at 500~nm wavelength), ground wind speed $v_\mathrm{g}$ = 5~m/s and zenith angle $\delta$ = 0~deg as a function of time. One can see that they are in close agreement with each other, taking into account their differing normalization factors. Here the Strehl ratio is defined as the ratio of peak on-axis intensity to that with no perturbations and does not take atmospheric transmission into account. Measured beacon power, on the other hand, is normalized to the power in a system with perfect alignment and no atmosphere. The average steady state Strehl is above 0.8, which is considered to be diffraction limited by the Maréchal criterion \cite{Born1975_marechal}. Fig. \ref{fig:spatialprofile} displays the time-averaged spatial profile and peak location as a function of time for the same beam. Spatial coordinates are in angular units to make them independent of propagation distance. The beam diameter is inversely proportional to array diameter, as the main lobe of a properly phased DEPA has the same geometric properties as a single aperture of the same size and shape \cite{Hettel2019_spie}. In this case the beam diameter is approximately 0.16~arcseconds (or 308~m for our propagation distance of $4\times10^8$~m). The peak location was found by performing a Gaussian fit around the maximum intensity value. The fitting procedure increases the resolution of the location values, which allows for subtle motion to be seen in the beam. See \textcolor{urlblue}{Visualization 1} for a video of the beam's spatial mode in these conditions, and \textcolor{urlblue}{Visualization 2} for a video in the harshest conditions simulated.

\begin{figure}
    \centering
    \includegraphics[width=0.7\linewidth]{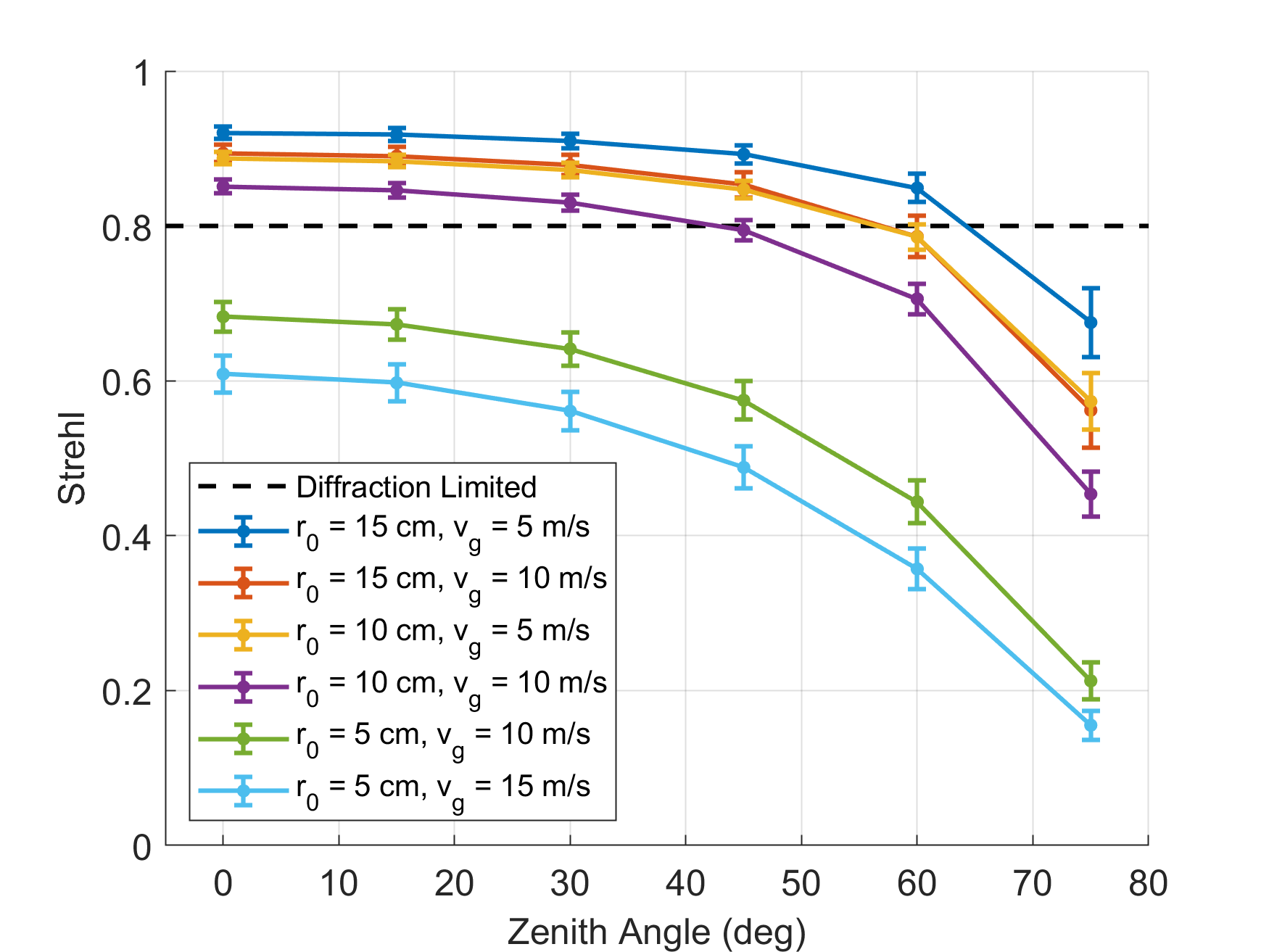}
    \caption{Strehl as a function of zenith angle for each atmospheric condition. Error bars represent the standard deviation over the 500~ms time series. The beam stays diffraction limited (Strehl > 0.8) for $r_0\geq10$~cm and $\delta\leq$ 30-60~deg depending on atmospheric conditions, where $r_0$ labels are specified for 500~nm wavelength and $\delta$ = 0~deg.}
\label{fig:strehlvszen}
\end{figure}

\begin{figure}[t!]
    \centering
    \includegraphics[width=0.7\linewidth]{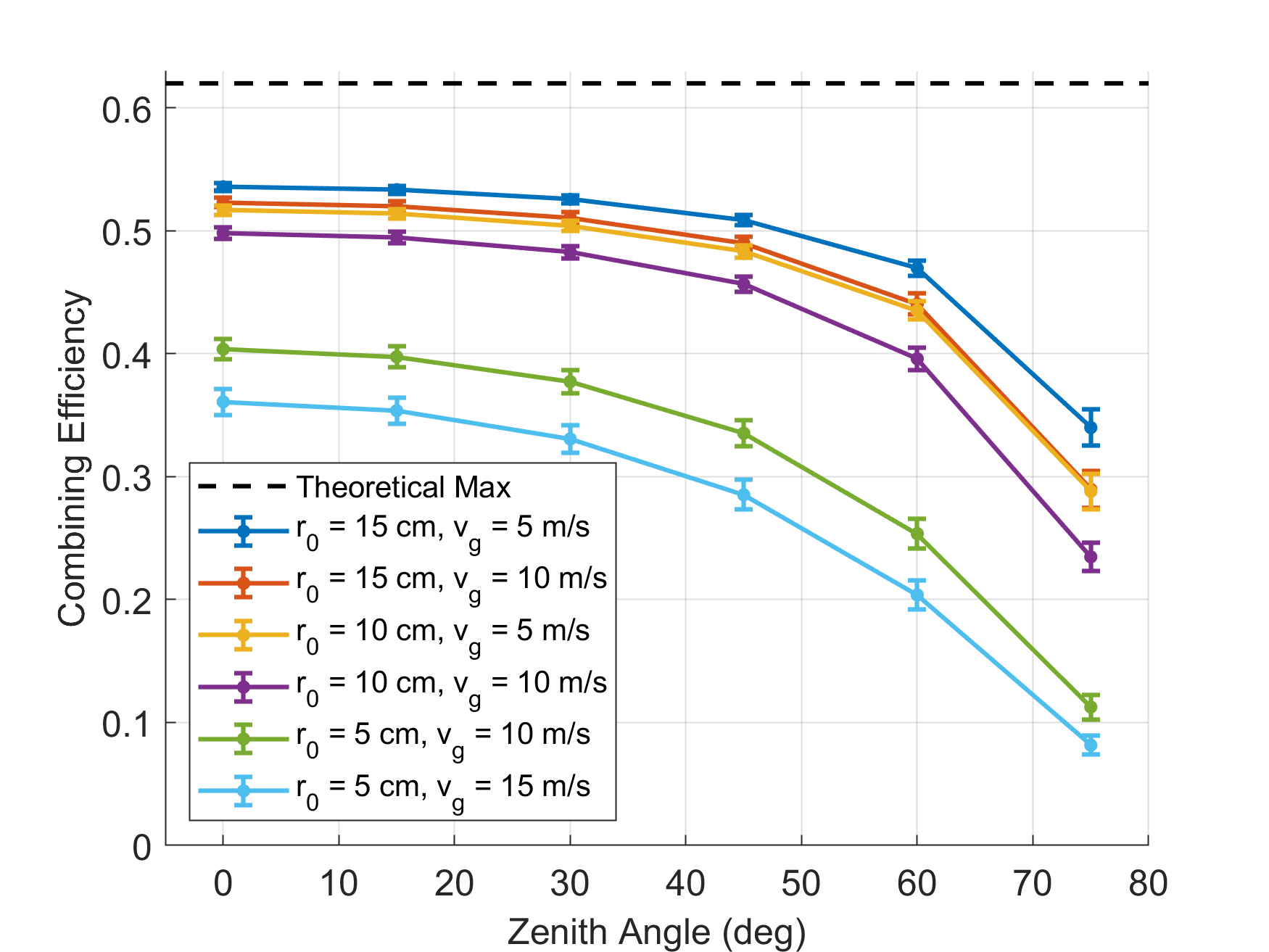}
    \caption{Combining efficiency as a function of zenith angle for each atmospheric condition. Error bars represent the standard deviation over the 500~ms time series. The dashed line is the theoretical maximum combining efficiency for an array composed of Gaussian beams transmitted through circular apertures. $r_0$ labels are specified for 500~nm wavelength and $\delta$ = 0~deg.}
    \label{fig:combeffvszen}
\end{figure}

Figs. \ref{fig:strehlvszen} and \ref{fig:combeffvszen} summarize Strehl and combining efficiency results for all of the simulations performed. Average Strehl and combining efficiency for each of the atmospheric models are displayed as a function of zenith angle. Note that for a given atmospheric model, $r_0$ decreases with zenith angle and the specified values of $r_0$ are for $\delta$ = 0~deg. The beam remained diffraction limited for $r_0$ down to 10~cm and $\delta$ up to 30-60~deg, depending on the model (see \textcolor{urlblue}{Visualization 3-4} for spatial mode). These conditions cover over 70\% of the site survey data taken for the TMT \cite{Schock2009_tmtdata}, a promising implication for high duty cycle applications of DEPAs at high altitude sites. 


\begin{figure}
    \centering
    \includegraphics[width=0.7\linewidth]{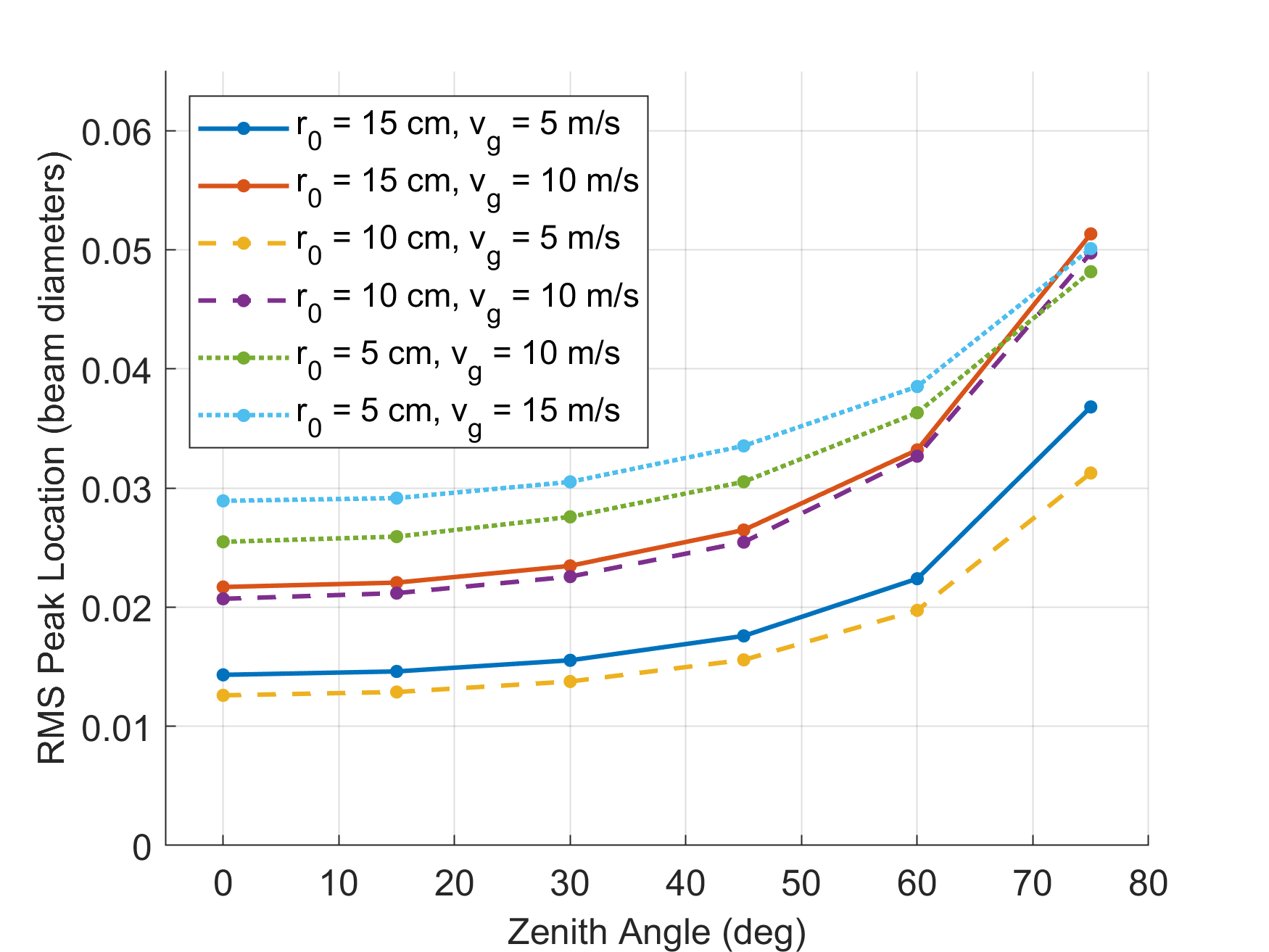}
    \caption{RMS peak location over 500~ms as a function of zenith angle for each atmospheric condition. The time series for $r_0$ = 15~cm, $v_\mathrm{g}$ = 5~m/s, and $\delta$ = 0~deg is plotted in Fig. \ref{fig:peaklocation}. $r_0$ labels are specified for 500~nm wavelength and $\delta$ = 0~deg.}
    \label{fig:rmspeak}
\end{figure}

The root-mean-square peak location as a function of zenith angle for each atmospheric condition is plotted in Fig. \ref{fig:rmspeak}. There is not a clear, intuitive relationship between $r_0$ and the magnitude of drift, as it is a function of both the strength and spatial structure of the residual phases. The beam drifts more as the linear component of the residual phase profile increases, which in some cases may be less prominent in higher turbulence than in lower turbulence. For a given atmospheric turbulence structure, however, the beam drifts more as both wind speed and zenith angle increase, as one might expect.

\begin{figure}
    \centering
    \includegraphics[width=0.7\textwidth]{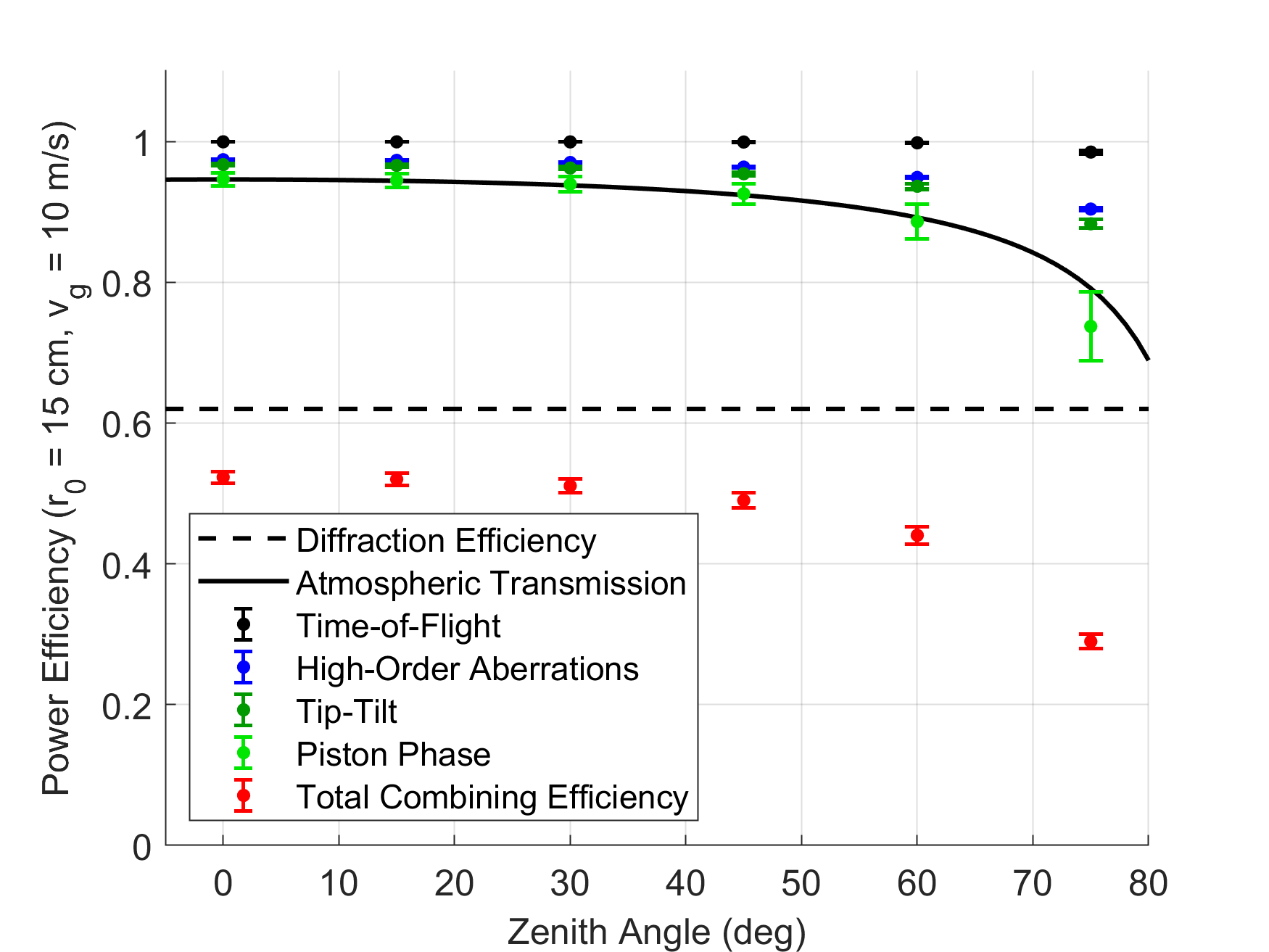}
    \caption{A breakdown of all of the loss mechanisms as a function of zenith angle in the simulations for $r_0$ = 15~cm and $v_\mathrm{g}$ = 10~m/s. Diffraction efficiency, atmospheric transmission, and time-of-flight (black) are all fundamental losses that cannot be avoided. High-order aberrations (blue) can only be mitigated by decreasing the sub-aperture diameter, which requires more subelements for a given array diameter. Piston phase and tip-tilt errors (green) may be reduced with ongoing hardware development. Other effects, such as those due to back reflections of the transmit beam or thermal blooming, are assumed to be sufficiently mitigated or negligible. Error bars represent the standard deviation over the 500~ms time series.}
\label{fig:lossmechanismsgood}
\end{figure}

It is instructive to analyze how each of the loss mechanisms in the simulation individually affect the DEPA's performance. Fig. \ref{fig:lossmechanismsgood} plots the contributing power efficiency loss factors for the $r_0$ = 15~cm, $v_\mathrm{g}$ = 10~m/s case as a function of zenith angle. Black items are fundamental losses which cannot be improved. High order aberrations, in blue, may be mitigated by decreasing the sub-aperture size. Piston phase and tip-tilt efficiency, in green, may be increased with further hardware and servomechanism development. Since the subelements are modular, all losses besides piston phase are applicable to any array size.

\section{Scaling to larger arrays}

With use of identical, modular subelements, it is expected that piston phase losses are the only thing that will change as a function of array size. Maintaining good phase integrity may be difficult for larger systems, as the closed loop bandwidth of SPGD is inversely proportional $N$. Long fiber baselines will also add phase noise \cite{Srinivasan2018_phasenoise}. However, since the proposed DEPA system can be built with nested splitting/combining junctions between the master oscillator and the sub-apertures, it is compatible with a nested SPGD configuration \cite{Goodno2014_nspgd}. Nesting may be accomplished by running SPGD in parallel with a detector at every junction, exponentially increasing the system's closed loop bandwidth for a given acquisition frequency and number of subelements. For example, if the subelements are nested in a 5-deep tree, over one billion can be phased together with only 64 sources at each junction. A demonstration that such a configuration can overcome the expected phase perturbations would suggest that kilometer scale systems placed at high altitude sites are capable of producing diffraction limited spots through the atmosphere.

\section{Conclusion}
Detailed dynamic beam propagation simulations of a directed energy phased array transmitting through atmosphere in conditions common to astronomical sites were performed using realistic system noise derived from laboratory measurements. A recently developed beam propagation code optimized via parallelization for very large numbers of elements (>$10^{10}$) in the array made the simulation of a various scenarios possible. The results provide evidence that large aperture ($>>r_0$) ground based systems at near-infrared wavelength are capable of producing a stable diffraction limited spot in outer space if the atmosphere has $r_0\geq$ 10~cm (at 500~nm wavelength and zenith angle = 0~deg) and zenith angles $\leq$ 60~deg. These conditions are often satisfied at current astronomical quality high altitude ground based sites. With the success of nested SPGD and the option to add an arbitrary number of subelements, such systems would overcome the size and power limits of current coherent directed energy technology and enable the delivery of high flux beams over vast distances. This development would enable transformational capabilities such as powering lunar bases or propelling spacecraft for rapid interplanetary travel, ablation of incoming asteroids, and even direct photon momentum transfer to enable the first relativistic spacecraft thus enabling interstellar flyby robotic missions \cite{Lubin2016_roadmap}. Similar technology may also be used for terrestrial applications such as powering of ground-based remote sensors and rovers in inhospitable locations, remote sensing LIDAR, laser machining, and matter in extreme EOS (equation of state) conditions.

\section*{Funding}
Funding for this program comes from NASA grants NIAC Phase I DEEP-IN – 2015 NNX15AL91G and NASA NIAC Phase II DEIS – 2016 NNX16AL32G and the NASA California Space Grant NASA NNX10AT93H as well as a generous gift from the Emmett and Gladys W. fund.

\section*{Acknowledgments}
PML acknowledges support from the Breakthrough Foundation as part of the Starshot program.

\section*{Disclosures}

The authors declare no conflicts of interest.

\section*{Data availability}

Data underlying the results presented in this paper are not publicly available at this time but may be obtained from the authors upon reasonable request.\\
\\
See \textcolor{urlblue}{Supplement 1} for supporting content.


\bibliography{references}

\begin{thebibliography}{10}
\newcommand{\enquote}[1]{``#1''}

\bibitem{Zervas2014_hplasers}
M.~N. {Zervas} and C.~A. {Codemard}, \enquote{High power fiber lasers: A
  review,} {\protect\JournalTitle{IEEE Journal of Selected Topics in Quantum
  Electronics}} \textbf{20}, 219--241 (2014).

\bibitem{Lubin2016_roadmap}
P.~Lubin, \enquote{A roadmap to interstellar flight,}
  {\protect\JournalTitle{Journal of the British Interplanetary Society - JBIS}}
  \textbf{69}, 40--72 (2016).

\bibitem{Lubin2020_ESA}
P.~Lubin and W.~Hettel, \enquote{The path to interstellar flight,}
  {\protect\JournalTitle{Acta Futura}} \textbf{12}, 9--44 (2020).

\bibitem{GeminiObs_transmission}
\enquote{The sites | gemini observatory,}
  \url{https://www.gemini.edu/observing/telescopes-and-sites/sites#0.9-2.7um}.

\bibitem{Hettel2019_spie}
W.~Hettel, P.~Meinhold, P.~Krogen, and P.~Lubin, \enquote{{Beam propagation
  simulation of large phased laser arrays},} in \emph{Laser Beam Shaping XIX,}
  vol. 11107 A.~Dudley and A.~V. Laskin, eds., International Society for Optics
  and Photonics (SPIE, 2019), pp. 153 -- 165.

\bibitem{Novotny2012_prop}
B.~L. Novotny and B.~Hecht, \enquote{Angular spectrum representation of optical
  fields,} in \emph{Principles of Nano-Optics,}  (Cambridge University Press,
  2012), pp. 39--40.

\bibitem{Reeves2016_soapy}
A.~Reeves, \enquote{{Soapy: an adaptive optics simulation written purely in
  Python for rapid concept development},} in \emph{Adaptive Optics Systems V,}
  vol. 9909 E.~Marchetti, L.~M. Close, and J.-P. Véran, eds., International
  Society for Optics and Photonics (SPIE, 2016), pp. 2173 -- 2183.

\bibitem{Schock2009_tmtdata}
M.~{Sch{\"o}ck}, S.~{Els}, R.~{Riddle}, W.~{Skidmore}, T.~{Travouillon},
  R.~{Blum}, E.~{Bustos}, G.~{Chanan}, S.~G. {Djorgovski}, P.~{Gillett},
  B.~{Gregory}, J.~{Nelson}, A.~{Ot{\'a}rola}, J.~{Seguel}, J.~{Vasquez},
  A.~{Walker}, D.~{Walker}, and L.~{Wang}, \enquote{{Thirty Meter Telescope
  Site Testing I: Overview},} {\protect\JournalTitle{Publ. Astr. Soc. Pac.}}
  \textbf{121}, 384 (2009).

\bibitem{cbc2013_spgd}
S.~Redmond, K.~Creedon, T.~Y. Fan, A.~Sanchez, C.~Yu, and J.~Donnelly,
  \enquote{Active coherent combination using hill climbing-based algorithms for
  fiber and semiconductor amplifiers,} in \emph{Coherent Laser Beam Combining,}
   (Wiley-VCH, 2013), pp. 103--136.

\bibitem{Andrews2001_scint}
L.~C. Andrews, R.~L. Phillips, and C.~Y. Young, \enquote{Modeling optical
  scintillation,} in \emph{Laser Beam Scintillation with Applications,}  vol.
  PM99 (SPIE Press, 2001), pp. 67--96.

\bibitem{Thomas2005_scint}
F.~E. Thomas, \enquote{The scintillation index in moderate to strong turbulence
  for the gaussian beam wave along a slant path,} Ph.D. thesis, University of
  Central Florida (2005).

\bibitem{Brooker1985_thinatm}
H.~G. {Booker}, J.~A. {Ferguson}, and H.~O. {Vats}, \enquote{{Comparison
  between the extended-medium and the phase-screen scintillation theories},}
  {\protect\JournalTitle{Journal of Atmospheric and Terrestrial Physics}}
  \textbf{47}, 381--399 (1985).

\bibitem{laserprop2005_thinatm}
L.~C. Andrews and R.~L. Phillips, \enquote{Propagation through random phase
  screens,} in \emph{Laser Beam Propagation through Random Media, Second
  Edition,}  (SPIE, 2005), pp. 648--666.

\bibitem{Born1975_marechal}
M.~Born and E.~Wolf, \emph{Principles of Optics, 5th ed.} (Pergamon Press,
  1975).

\bibitem{cbc2013_combining}
A.~Brignon, J.~Bourderionnet, C.~Bellanger, and J.~Primot, \enquote{Collective
  techniques for coherent beam combining of fiber amplifiers,} in
  \emph{Coherent Laser Beam Combining,}  (Wiley-VCH, 2013), pp. 137--165.

\bibitem{Srinivasan2018_phasenoise}
P.~Srinivasan, P.~Krogen, P.~Meinhold, W.~Hettel, O.~Ou, P.~M. Lubin,
  N.~Blasey, and G.~B. Hughes, \enquote{{Beamed-energy propulsion: optical
  phase noise in 1064nm fiber amplifiers },} in \emph{Optical Modeling and
  Performance Predictions X,}  vol. 10743 M.~A. Kahan and M.~B. Levine-West,
  eds., International Society for Optics and Photonics (SPIE, 2018), pp. 99 --
  108.

\bibitem{Goodno2014_nspgd}
G.~D. Goodno, E.~Cheung, and W.~H. Long, \enquote{Nested loop coherent beam
  combining system,}  (2014). US8837033B2.

\end{thebibliography}


\begin{thebibliography}{10}
\newcommand{\enquote}[1]{``#1''}

\bibitem{cbc2013_spgd}
S.~Redmond, K.~Creedon, T.~Y. Fan, A.~Sanchez, C.~Yu, and J.~Donnelly,
  \enquote{Active coherent combination using hill climbing-based algorithms for
  fiber and semiconductor amplifiers,} in \emph{Coherent Laser Beam Combining,}
   (Wiley-VCH, 2013), pp. 103--136.

\bibitem{Srinivasan2018_phasenoise}
P.~Srinivasan, P.~Krogen, P.~Meinhold, W.~Hettel, O.~Ou, P.~M. Lubin,
  N.~Blasey, and G.~B. Hughes, \enquote{{Beamed-energy propulsion: optical
  phase noise in 1064nm fiber amplifiers },} in \emph{Optical Modeling and
  Performance Predictions X,}  vol. 10743 M.~A. Kahan and M.~B. Levine-West,
  eds., International Society for Optics and Photonics (SPIE, 2018), pp. 99 --
  108.

\bibitem{Hettel2020_actuator}
W.~Hettel, B.~Phillips, A.~N. Cohen, P.~Meinhold, P.~Krogen, P.~Srinivasan,
  J.~Erlikhman, J.~Meinhold, and P.~Lubin, \enquote{A compact voice coil driven
  optical tracking system for directed energy phased arrays,} Manuscript
  submitted for publication.

\bibitem{Lubin2020_ESA}
P.~Lubin and W.~Hettel, \enquote{The path to interstellar flight,}
  {\protect\JournalTitle{Acta Futura}} \textbf{12}, 9--44 (2020).

\bibitem{Mahajan1983_marechal}
V.~N. Mahajan, \enquote{Strehl ratio for primary aberrations in terms of their
  aberration variance,} {\protect\JournalTitle{J. Opt. Soc. Am.}} \textbf{73},
  860--861 (1983).

\bibitem{Ross2009_marechal}
T.~S. Ross, \enquote{Limitations and applicability of the mar\'{e}chal
  approximation,} {\protect\JournalTitle{Appl. Opt.}} \textbf{48}, 1812--1818
  (2009).

\bibitem{Noll1976_zernike}
R.~J. Noll, \enquote{Zernike polynomials and atmospheric turbulence,}
  {\protect\JournalTitle{J. Opt. Soc. Am.}} \textbf{66}, 207--211 (1976).

\bibitem{Tyson2004_aoguide}
R.~K. Tyson and B.~W. Frazier, \enquote{Field guide to adaptive optics,} in
  \emph{SPIE Field Guides,}  vol. FG03 (SPIE Press, 2004).

\bibitem{Toyoshima2011_HVmodel}
M.~Toyoshima, H.~Takenaka, and Y.~Takayama, \enquote{Atmospheric
  turbulence-induced fading channel model for space-to-ground laser
  communications links,} {\protect\JournalTitle{Opt. Express}} \textbf{19},
  15965--15975 (2011).

\bibitem{Mohr2010_windmodel}
J.~Mohr, R.~Johnston, and P.~Cottrell, \enquote{Optical turbulence measurements
  and models for mount john university observatory,}
  {\protect\JournalTitle{Publications of the Astronomical Society of
  Australia}} \textbf{27} (2010).

\bibitem{Schock2009_tmtdata}
M.~{Sch{\"o}ck}, S.~{Els}, R.~{Riddle}, W.~{Skidmore}, T.~{Travouillon},
  R.~{Blum}, E.~{Bustos}, G.~{Chanan}, S.~G. {Djorgovski}, P.~{Gillett},
  B.~{Gregory}, J.~{Nelson}, A.~{Ot{\'a}rola}, J.~{Seguel}, J.~{Vasquez},
  A.~{Walker}, D.~{Walker}, and L.~{Wang}, \enquote{{Thirty Meter Telescope
  Site Testing I: Overview},} {\protect\JournalTitle{Publ. Astr. Soc. Pac.}}
  \textbf{121}, 384 (2009).

\bibitem{Reeves2016_soapy}
A.~Reeves, \enquote{{Soapy: an adaptive optics simulation written purely in
  Python for rapid concept development},} in \emph{Adaptive Optics Systems V,}
  vol. 9909 E.~Marchetti, L.~M. Close, and J.-P. Véran, eds., International
  Society for Optics and Photonics (SPIE, 2016), pp. 2173 -- 2183.

\bibitem{Townson2019_aotools}
M.~J. Townson, O.~J.~D. Farley, G.~O. de~Xivry, J.~Osborn, and A.~P. Reeves,
  \enquote{Aotools: a python package for adaptive optics modelling and
  analysis,} {\protect\JournalTitle{Opt. Express}} \textbf{27}, 31316--31329
  (2019).

\end{thebibliography}






\end{document}


\maketitle
\begin{figure}
    \centering
    \includegraphics[width=1\textwidth]{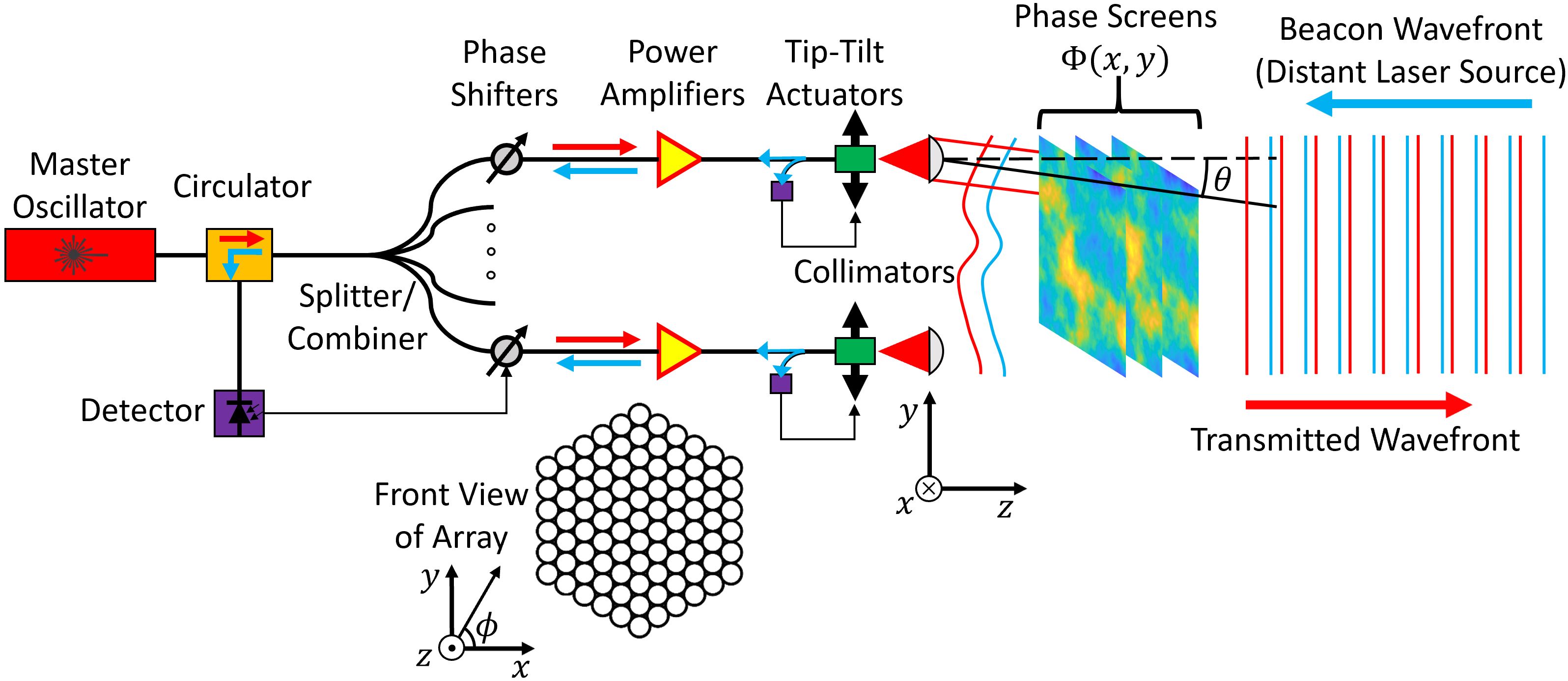}
    \caption{Schematic diagram of the simulated system with coordinates. A beacon laser signal (right, blue) passes through atmospheric phase screens and illuminates the DEPA (left), which measures the reverse-propagating beacon in fiber for phase conjugation and alignment feedback. The transmit beam (red) originates from a master seed laser, which is split and amplified in fiber before propagating in free space. Here a 91 element hexagonal array is shown, but the DEPA can be scaled to any size or shape.}
\label{fig:schematic}
\end{figure}

\section{Modeling of perturbations and adaptive optics}
\label{sec:perturbations}
Each subelement of the DEPA system is assumed to have a phase shifter to control piston phase (average phase over sub-aperture) and tip-tilt actuators to control output angle. It is also assumed that the DEPA is in a bidirectional master oscillator power amplifier (MOPA) configuration such that the beacon light can enter the system and be measured after combining at a coherent junction (see Fig. \ref{fig:schematic}). The bidirectional optical amplifier will be described in an upcoming paper. In an ideal case with no atmosphere (space environment), the inherent optical reciprocity of this bidirectional system ensures that optimizing coherently combined  beacon power at the location of the master oscillator also optimizes the power of the transmit beam at the source of the beacon. With an atmosphere (earth based array), scintillation and time-of-flight effects break this reciprocity, but the principle is the same.

\subsection{Piston phase}
\label{subsec:pistonphase}
SPGD is routinely used in adaptive optics (AO) systems where measurement of the individual sub-aperture phases is unreasonable. In this method, a beacon or laser guide star provides the wavefront reference. Elements of the AO system are dithered to measure the local wavefront gradient and adjusted actively to compensate for it. The control equation is given by \cite{cbc2013_spgd}:
\begin{equation}
    A_{i,k+1} = g_\mathrm{leakage}A_{i,k} + g\left( \frac{1}{2\sigma_\mathrm{rms}^2}\right) \left( \frac{S_+ - S_-}{S_+ + S_-} \right)D_{i,k}.
\end{equation}
Here $A_{i,k}$ represents the correction phase applied to element $i$ for iteration $k$, $D_{i,k}$ is its dither value, $S_+$ and $S_-$ are measurements of the performance metric for positive and negative dithers, $\sigma_\mathrm{rms}$ is the root mean square amplitude of $D$, and $g$ and $g_\mathrm{leakage}$ are gain terms. Our simulations use a Hadamard matrix for $D$, $\sigma_\mathrm{rms}$ = 0.01~waves ($\pi/50$~radians), $g_\mathrm{leakage}$ = 1, $g$ = 2, and an iteration frequency $f_\mathrm{i}$ = 1~MHz. $S_+$ and $S_-$ are computed by squaring the amplitude of the sum of the phase shifted electric fields collected by each subelement (see subsection \ref{subsec:power}) and represent the combined beacon power at the detector in Fig \ref{fig:schematic}. The phase convergence time $t_\mathrm{c}$ is
\begin{equation}
\label{eq:convergence}
    t_\mathrm{c} = \frac{\alpha N}{f_\mathrm{i}}
\end{equation}
for $N$ elements. $\alpha$ is a system-dependent coefficient and represents how many cycles through the dither matrix it takes for the performance metric to reach a steady state. With an orthonormal dither matrix (such as a Hadamard) it is typical to have $\alpha$$\sim$2 \cite{cbc2013_spgd}.

Atmospheric phase inputs are generated by spatially averaging the simulated beacon phase over the area occupied by each sub-aperture. These phases are added to randomly generated phases based on measured optical phase noise power spectra in fiber amplifiers \cite{Srinivasan2018_phasenoise}. Detector noise is generated with a flat power spectrum, and scaled to have a noise-equivalent power (NEP) 40~dB below the nominal beacon power per subelement in order to provide a sufficient signal-to-noise ratio. For example, if the nominal beacon power per subelement were 1~nW, this corresponds to a NEP of $10^{-13} \ \mathrm{W}/\sqrt{\mathrm{Hz}}$, which is a readily available specification from commercial sources. The residual phase errors of the SPGD simulation are then used as perturbations for the transmit beam.

\subsection{Tip-tilt}
\label{subsec:tiptilt}
The slope of the phase over an aperture will cause the beam to ``wander,'' as a phase perturbation that is linear with position is equivalent to an angular displacement of the beam in the focal plane. If the subelements of the DEPA are large fiber collimators, this angular displacement may be compensated for by moving the fiber in the focal plane of the lens. Under the paraxial approximation these displacements are calculated as:
\begin{equation}
\label{eq:tiptilt}
    \theta\cos{\phi}\boldsymbol{\hat{x}} + \theta\sin{\phi}\boldsymbol{\hat{y}} = \nabla\Phi(x,y)/k
\end{equation}
\begin{wrapfigure}{r}{0.45\linewidth}
\includegraphics[width=1\linewidth]{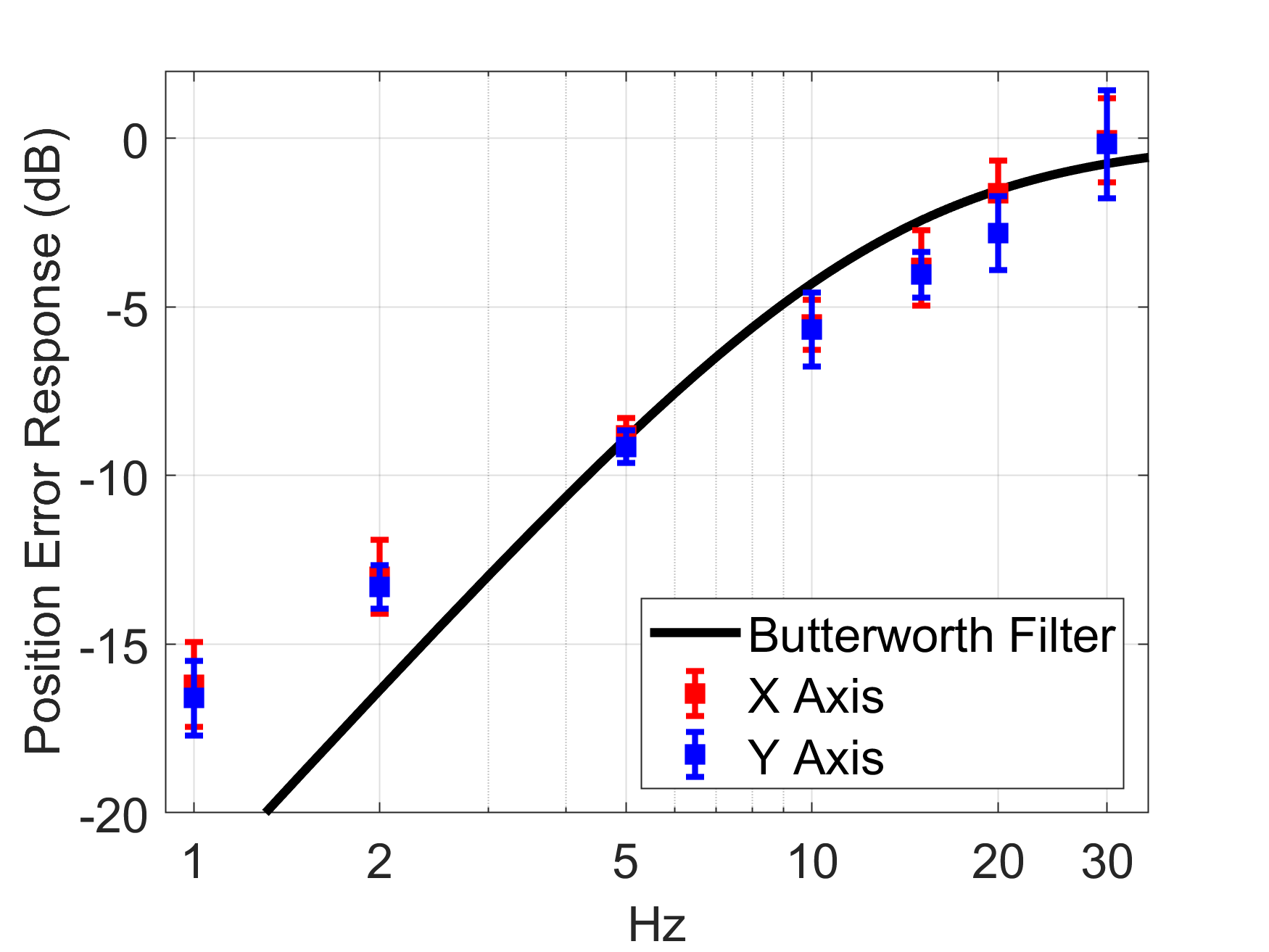}
\caption{Servo rejection of tip-tilt actuators in development at University of California, Santa Barbara. The residual position error can be approximated as the setpoint passed through a first order Butterworth high pass filter with $f_\mathrm{c}$ = 13~Hz. For these simulations, we assume we reach our goal of $f_\mathrm{c}$ = 40~Hz.}
\label{fig:tiptilt}
\end{wrapfigure}
where $\theta$ is the angle from the target trajectory, $\phi$ is the azimuthal angle, $\nabla\Phi(x,y)$ is the gradient of the atmosphere-induced optical phase profile, $k$ is the wave number and $x$ and $y$ are in the horizontal plane (see Fig. \ref{fig:schematic}). The two components of $\theta$ are ``tip'' and ``tilt,'' respectively. Although Eq. (\ref{eq:tiptilt}) is for a point as opposed to a finite aperture, the finite aperture values are approximated by spatially averaging $\nabla\Phi(x,y)$ over the area occupied by each sub-aperture.

The Experimental Cosmology Group at University of California, Santa Barbara has recently developed compact and easily manufacturable actuators for tip-tilt correction \cite{Hettel2020_actuator}. These actuators use light received from the beacon for feedback by modulating the fiber position in a circular perturbation at multiple kHz frequencies and using lock-in amplification to produce a two-dimensional error signal. Experimental data shows that this system roughly acts as a first order Butterworth high pass filter for positional error with a cutoff frequency of $f_\mathrm{c}$ = 13~Hz (see Fig. \ref{fig:tiptilt}). With further design optimization we expect to bring $f_\mathrm{c}$ up to 40~Hz. To model our target actuator performance, the simulated tip and tilt time series are passed through such a filter. A 4.4~kHz tone with an amplitude of 150~nrad is added to simulate the modulation used by the servo. The amplitude of the modulation is about 0.01 of the full angle divergence of a 10~cm diameter optic at 1.064~$\mu$m wavelength, well above what was used in \cite{Hettel2020_actuator}.

\subsection{Received and transmitted power}
\label{subsec:power}
The beacon power received by each subelement $P_\mathrm{beac}$ is calculated as inputs for the SPGD simulation. First, the intensity values from the beacon propagation step ($I_\mathrm{beac}$) are scaled by the atmospheric transmission and summed over the area of the sub-aperture to find the total amount of power incident on it. The atmospheric transmission $T$ is a function of zenith angle $\delta$. For a simple slab model,

\begin{equation}
    T = 1 - \sec{\delta}(1-T_0)
\end{equation}
where $T_0$ is the transmission for $\delta = 0$. We take $T_0 = 0.946$ at 1.064~$\mu$m wavelength based on Modtran simulations for a site at 4~km altitude \cite{Lubin2020_ESA}.

The amount of power coupled into the fiber is then approximated as the power incident on the aperture multiplied by the normalized on-axis intensity in the focal plane of the lens. The normalized on-axis intensity is found by combining losses from tip-tilt error and residual high-order phase aberrations over the sub-aperture (exponential terms in Eq. (\ref{eq:beaconpower})). The residual phase $\Phi_r(x,y)$ is calculated by removing the tip and tilt components of $\Phi(x,y)$:
\begin{eqnarray}
\Phi_r(x,y) &=& \Phi(x,y) - \overline{\nabla\Phi_x} \cdot x - \overline{\nabla\Phi_y} \cdot y
\end{eqnarray}
where $\overline{\nabla\Phi_x}$ and $\overline{\nabla\Phi_y}$ are the $x$ and $y$ components of $\nabla\Phi(x,y)$ spatially averaged over the sub-aperture. The variance of the residual phase, which we will denote as $\sigma_\Phi^2$, is then plugged into the Maréchal approximation \cite{Mahajan1983_marechal,Ross2009_marechal}, and we have
\begin{equation}
    P_\mathrm{beac} = Te^{-2(\theta/\theta_{1/2})^2}e^{-\sigma_\Phi^2}\sum_\mathrm{aperture} I_\mathrm{beac}\Delta x \Delta y
\label{eq:beaconpower}
\end{equation}
where $\theta_{1/2}$ is the half angle divergence of the subelement beam.

The power transmitted by each subelement is scaled by $T$ and otherwise assumed not to change, although the far field intensity reduction due to the high-order phase variance $\sigma_\Phi^2$ is accounted for by defocusing the beam by the appropriate amount. This approximation conserves the amount of transmitted power and is a reasonable assumption, since most of the residual phase variance is in the form of focusing perturbations if a Kolmogorov turbulence spectrum is assumed \cite{Noll1976_zernike}.

\subsection{Time-of-Flight}
\label{subsec:TOF}
The finite speed of light becomes a factor in DEPA performance as the atmosphere changes in the time between the beacon and transmit beams passing through it. This effect is accounted for by advancing each phase screen at height $h$ by the time-of-flight delay $\Delta t_\mathrm{TOF}$ in a slab model:
\begin{equation}
    \Delta t_\mathrm{TOF} = 2h\sec{\delta}/c
\end{equation}
where $c$ is the speed of light and $\delta$ is the angle from zenith. These phase screens are added together and subtracted from $\Phi(x,y)$. The differential is then used to produce residual errors for the transmit beam in the same manner as described in subsections \ref{subsec:pistonphase}-\ref{subsec:power}. This calculation again assumes that the atmosphere is thin compared to the target distance.

\section{Atmospheric modeling and phase screen generation}
\label{sec:atmosphere}
\subsection{Turbulence model}
The Hufnagel-Valley (HV) model is commonly used to describe the atmospheric turbulence structure function $C_n^2$ as a function of altitude \cite{Tyson2004_aoguide}. It includes three terms, which respectively correspond to a turbulence peak in the tropopause layer, an exponential decay through the troposphere, and a strong turbulence patch near the ground:
\begin{equation}
\label{eq:HV}
    C_n^2(h) = 5.94 \times 10^{-23}h^{10}\left(\frac{v_\mathrm{rms}}{27}\right)^2e^{-h} + 2.7 \times 10^{-16}e^{-2h/3} + Ae^{-10h}
\end{equation}
where $v_\mathrm{rms}$ is the root mean square wind velocity in meters per second \cite{Toyoshima2011_HVmodel}, $h$ is altitude above ground level (AGL) in kilometers, and $A$ is a tunable parameter to set the strength of turbulence at the surface. $C_n^2$ has units of $\mathrm{m}^{-2/3}$. A common condition of choice is the HV 5-7 model, where $v_\mathrm{rms}=21$~m/s and $A=1.7 \times 10^{-14}~\mathrm{m}^{-2/3}$ to yield a Fried parameter $r_0$ of 5~cm. The Fried parameter, which sets the size scale of the total integrated turbulence, is related to $C_n^2$ by
\begin{equation}
\label{eq:r0}
    r_0 = \left[0.423k^2\sec{\delta}\int_{h_1}^{h_2} C_n^2(h)dh \right]^{-3/5}
\end{equation}
for a plane wave of wave number $k$ which propagates along a path from $h_1$ to $h_2$ with a zenith angle $\delta$. Since $r_0$ is wavelength dependent, it is typical for astronomers to specify $r_0$ at 500~nm wavelength. Although $r_0$ is also dependent on $\delta$, for simplicity we specify its value at $\delta = 0$ and also specify $\delta$. The tuning parameters $A$ and $v_\mathrm{rms}$ alone unfortunately do not provide enough control over $r_0$ to simulate some low turbulence conditions of interest to us. For example, even if we set $A=0$ and $v_\mathrm{rms} = 21$~m/s we obtain $r_0$ $\sim$11~cm for 500~nm wavelength, which prevents the simulation of some common conditions at high altitude sites (see Fig. \ref{fig:wind&r0overlay}). To provide more tunability, we set $A=0$ and introduce a parameter $B$ to modify the troposphere's exponential decay term:
\begin{equation}
\label{eq:HVmod}
    C_n^2(h) = 5.94 \times 10^{-23}h^{10}\left(\frac{v_\mathrm{rms}}{27}\right)^2e^{-h} + B \times 2.7 \times 10^{-16}e^{-2h/3}.
\end{equation}
This modification is conservative in how it affects the DEPA performance for a given value of $r_0$. If we were to compare two atmospheres described by Eqs. (\ref{eq:HV}) and (\ref{eq:HVmod}) with the same $r_0$ and $A>0$, by Eq. (\ref{eq:r0}) they would have the same total integrated turbulence along the path. Therefore, in Eq. (\ref{eq:HVmod}) the turbulence that is lost from the surface patch is simply redistributed to higher altitudes. The shape of the $C_n^2$ distribution primarily affects the degree at which the beacon is scintillated and the time scale at which the perturbations change. Beacon scintillation increases as turbulence is redistributed to be further away from the DEPA, as this is a result of light propagating after it hits turbulent patches. The time scale of the perturbations is dominated by wind speed, which typically increases with altitude and reaches its maximum at the tropopause layer (see subsection \ref{sec:windmodel}). Both increasing beacon scintillation and the time scale of perturbations will worsen the DEPA performance. Thus, by ignoring ground effects and introducing $B$, we do not ease the performance requirements of the DEPA from those that would be determined by the use of a standard HV model with the same $r_0$.

\subsection{Wind model}
\label{sec:windmodel}
Wind speed as a function of altitude is commonly expressed using a Greenwood wind model, which assumes a Gaussian profile with the peak at the tropopause layer \cite{Mohr2010_windmodel,Tyson2004_aoguide}. The Bufton wind model \cite{Tyson2004_aoguide} is a specific Greenwood model where the optical system is pointed at zenith, the ground wind speed is 5~m/s, and the tropopause layer is 9.4~km AGL (Above Ground Level) with a width parameter of 4.8~km and peak velocity of 30~m/s. This study uses a modified Bufton wind model with a tunable ground wind speed $v_\mathrm{g}$:
\begin{equation}
\label{eq:bufton}
    V(h) = v_\mathrm{g} + 30\mathrm{exp}\left[ -\left(\frac{h-9.4}{4.8}\right)^2\right]
\end{equation}
which allows $v_\mathrm{rms}$ to be computed as
\begin{equation}
    v_\mathrm{rms} = \left(\frac{1}{h_2-h_1}\int_{h_1}^{h_2} V^2(h)dh \right)^{1/2}
\end{equation}
which is then plugged into Eq. (\ref{eq:HVmod}). When this model is used for $\delta > 0$, it assumes a worst case scenario of the wind being perpendicular to the propagation direction. Under these definitions, the structure and dynamics of the atmosphere are completely determined by $r_0$ and $v_\mathrm{g}$.

\subsection{Choice of atmospheric conditions}
\begin{figure}
    \centering
        \begin{subfigure}[b]{0.45\linewidth}
        \includegraphics[width=\textwidth]{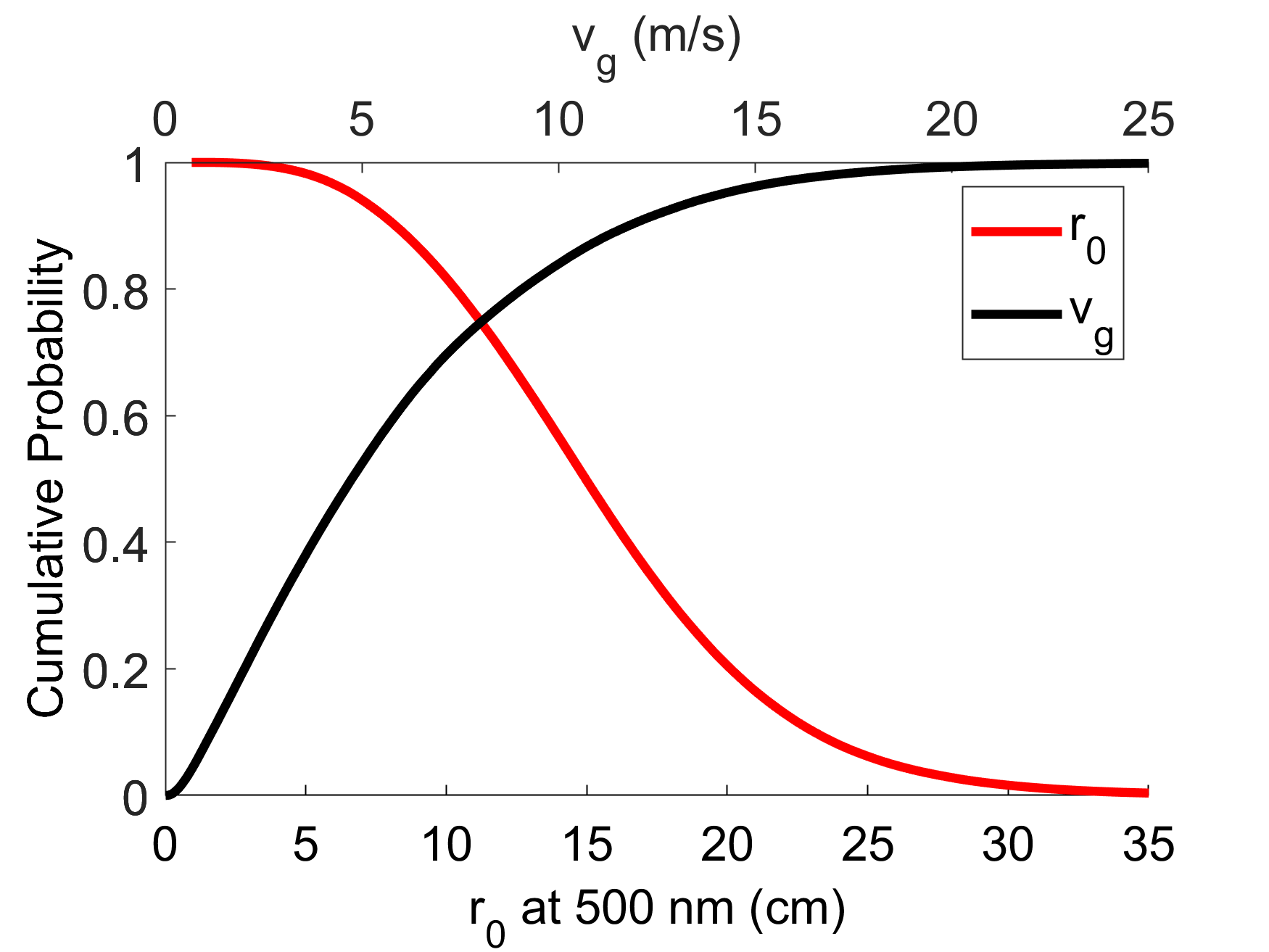}
        \caption{Cumulative probabilities for ground wind speed ($v_\mathrm{g}$) and Fried parameter ($r_0$) based on TMT site survey data.}
        \label{fig:wind&r0cumulative}
    \end{subfigure}
    ~
    \begin{subfigure}[b]{0.45\linewidth}
        \includegraphics[width=\linewidth]{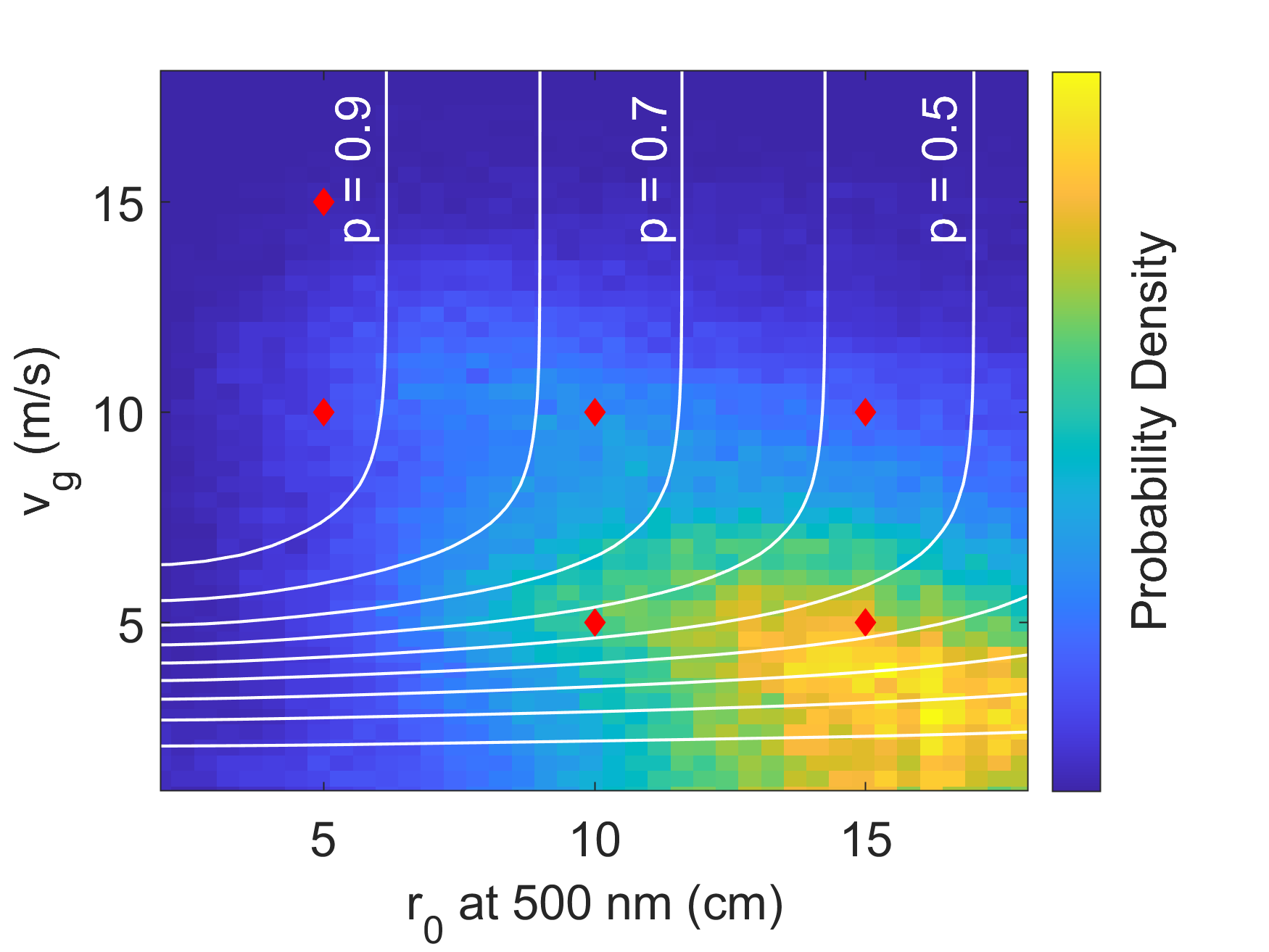}
        \caption{Probability density function of $r_0$ and $v_\mathrm{g}$. Red diamonds show simulated conditions, and white contours show cumulative probability $p$ within the data set.}
        \label{fig:wind&r0overlay}
    \end{subfigure}

    \caption{Fig. \ref{fig:wind&r0cumulative} shows the probability of having $r_0$ above or $v_\mathrm{g}$ below the specified value based on site survey data taken for the Thirty Meter Telescope (TMT). $r_0$ values are for 500~nm wavelength. Fig. \ref{fig:wind&r0overlay} shows the probability density for a given combination of $r_0$ and $v_\mathrm{g}$ from the same data. It should be noted that some high wind speed data points were discarded for this plot because $r_0$ was not measured at these times. Red diamonds mark the conditions simulated in this paper. The white contours show the cumulative probability $p$ of achieving a given set of conditions or better, and have a spacing of 0.1. The cumulative probability in the top left corner is 1.}
\label{fig:tmtdata}
\end{figure}

Data collected at proposed sites for the Thirty Meter Telescope (TMT) \cite{Schock2009_tmtdata} were used to determine appropriate values of $r_0$ and $v_\mathrm{g}$. These sites include Cerro Tolar, Cerro Armazones, Cerro, Tolonchar, San Pedro Mártir, and Mauna Kea. Probability functions and cumulative distributions were produced from DIMM seeing and sonic anemometer data, shown in Fig. \ref{fig:tmtdata}. $r_0$ values are for 500~nm wavelength. Our chosen conditions to simulate are denoted by the red diamonds on Fig. \ref{fig:wind&r0overlay}. These points, which include $r_0$ = 5-15~cm and $v_\mathrm{g}$ = 5-15~m/s, were picked to cover some of the most probable conditions expected on a clear night at an astronomical site as well as a variety of poor conditions to evaluate the robustness of the system.

\begin{figure}
    \centering
        \begin{subfigure}[b]{0.45\linewidth}
        \includegraphics[width=\textwidth]{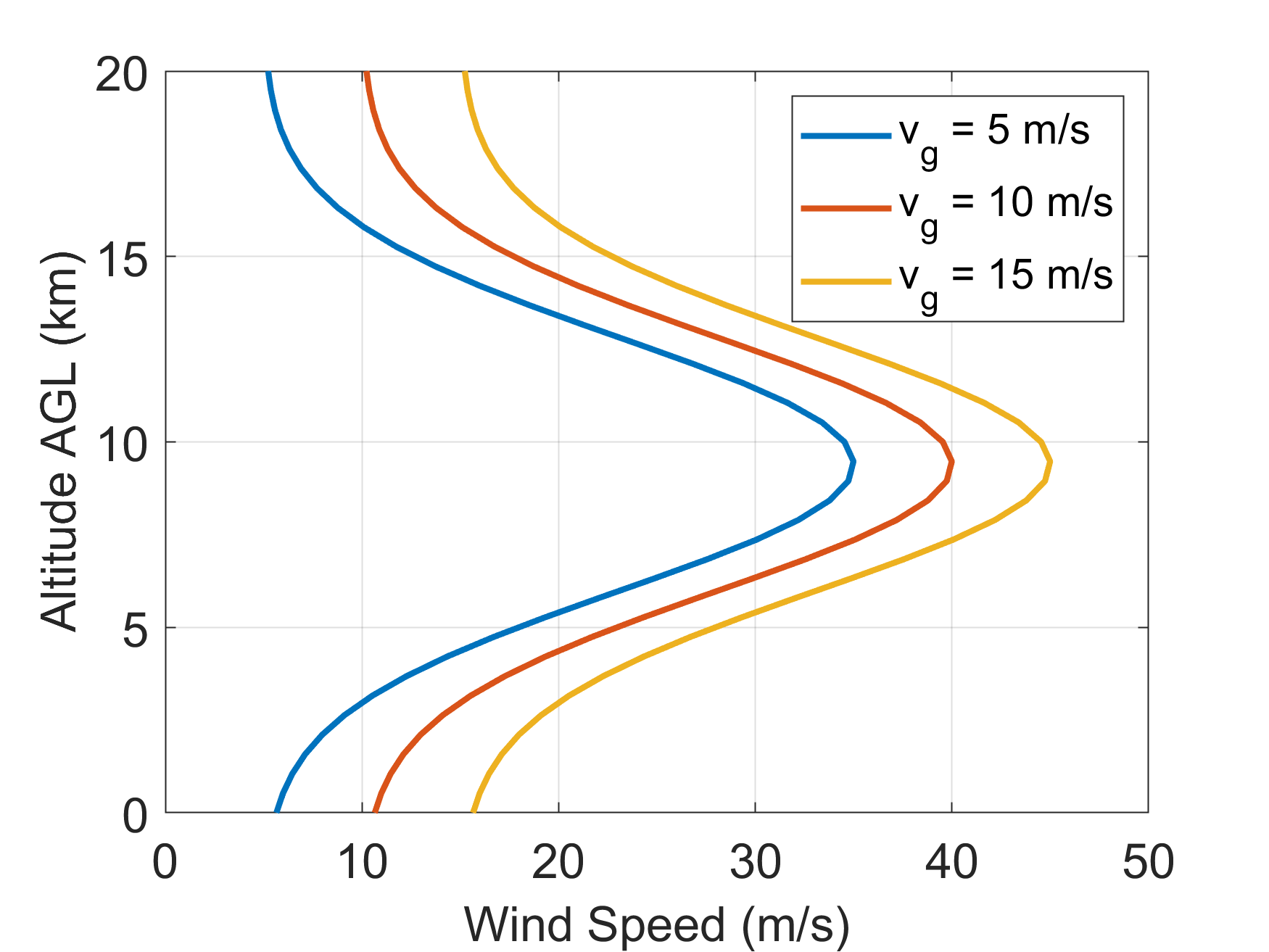}
        \caption{Wind speed as a function of altitude above ground level for each of the simulated conditions.}
        \label{fig:windinputs}
    \end{subfigure}
    ~
    \begin{subfigure}[b]{0.45\linewidth}
        \includegraphics[width=\linewidth]{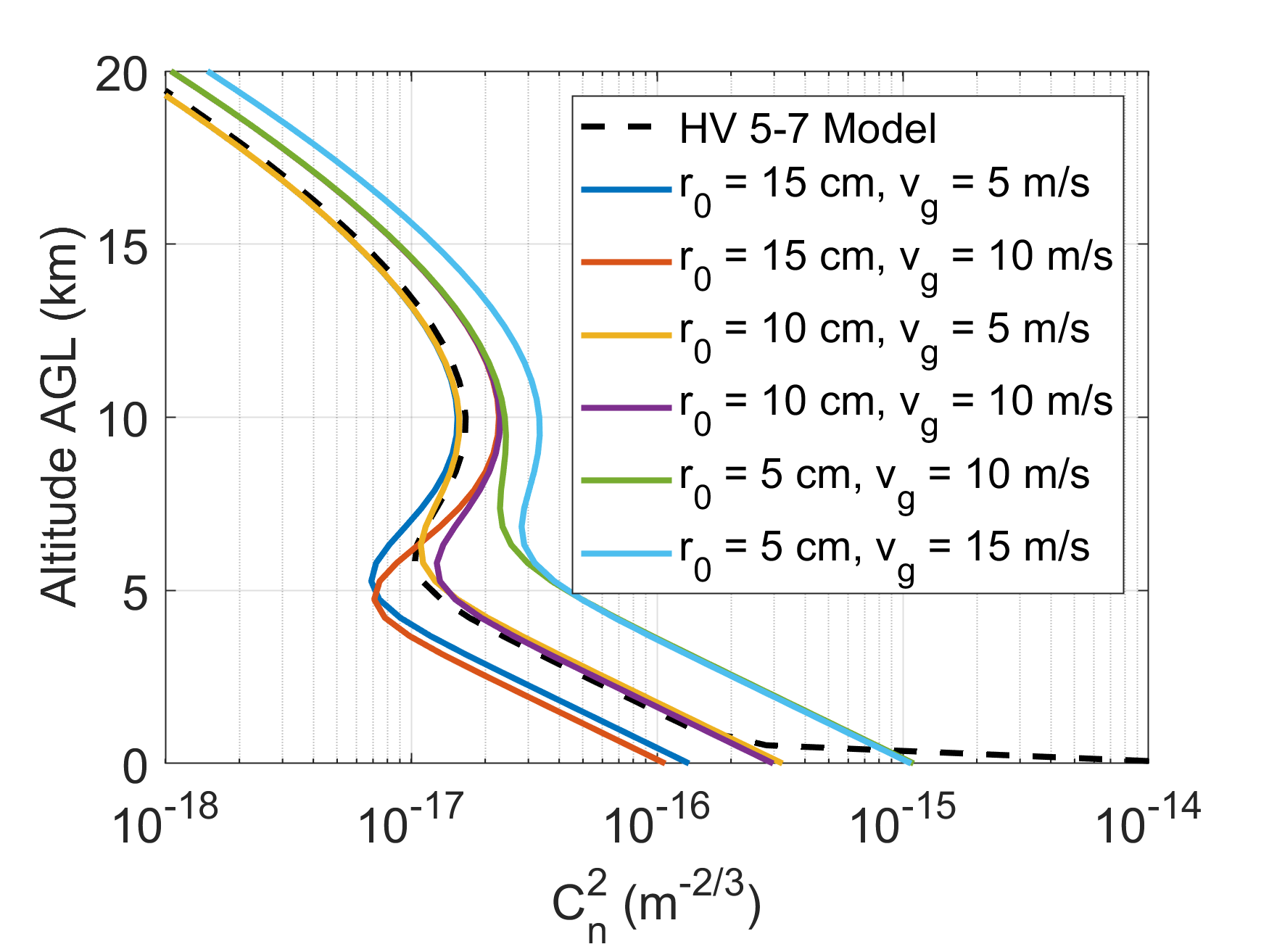}
        \caption{$C_n^2$ as a function of altitude above ground level for each of the simulated conditions. The HV 5-7 model is also shown for reference.}
        \label{fig:cn2inputs}
    \end{subfigure}

    \caption{Wind speed (left) and $C_n^2$ (right) profiles for the simulated atmospheric conditions, based on Eqs. (\ref{eq:HVmod}) and (\ref{eq:bufton}). The HV 5-7 model is shown for reference in Fig. \ref{fig:cn2inputs}. $r_0$ values are for 500~nm wavelength and $\delta$ = 0~deg.}
\label{fig:inputmodels}
\end{figure}

\subsection{Phase screen generation}
Atmospheric phase screens were produced using SOAPY \cite{Reeves2016_soapy}, a Python extension for modeling adaptive optical systems based on AOtools \cite{Townson2019_aotools}. It produces a specified number of two dimensional phase screens that correspond to turbulence patches at different heights based on a Kolmogorov spectrum. SOAPY allows the user to input custom wind and $C_n^2$ profiles, which we computed using Eqs. (\ref{eq:HVmod}) and (\ref{eq:bufton}) and are shown in Fig. \ref{fig:inputmodels}. For our desired fidelity, we generated phase screens with 2~cm horizontal resolution, 526~m vertical resolution, and 100~$\mu$s temporal resolution. The height of each atmosphere was 20~km with 0.5~s duration. Increasing the temporal or spatial resolution did not change the results to the reported accuracy. For angles off of zenith ($\delta > 0$), the distance between screens was scaled by $\sec{\delta}$ and the phase values by $\sqrt{\sec{\delta}}$ to preserve $C_n^2$. Since the SPGD simulation requires higher time resolution, intermediate phase and power inputs from Section \ref{sec:perturbations} were generated via spline interpolation and the phase outputs were downsampled back to the original frequency.

\bibliography{references}
